%% file: WeakCouplingSpectrum2008.tex
\documentclass[11pt]{article}

\usepackage{amsmath,amssymb}
\usepackage{graphicx}
\usepackage{stmaryrd}
\usepackage{overpic}
\usepackage{hyperref}

\makeatletter

\renewcommand{\section}{\@startsection{section}{1}{\z@}
                                   {-3.5ex \@plus -1ex \@minus -.2ex}
                                   {2.3ex \@plus .2ex}
                                   {\normalfont\large\bfseries}}
\renewcommand{\subsection}{\@startsection{subsection}{2}{\z@}
                                   {-3.25ex\@plus -1ex \@minus -.2ex}
                                   {1.5ex \@plus .2ex}
                                   {\normalfont\normalsize\bfseries}}
\renewcommand{\subsubsection}{\@startsection{subsubsection}{3}{\z@}
                                   {-3.25ex\@plus -1ex \@minus -.2ex}
                                   {1.5ex \@plus .2ex}
                                   {\normalfont\normalsize\bfseries}}
\renewcommand{\paragraph}{\@startsection{paragraph}{4}{\z@}
                                   {3.25ex \@plus1ex \@minus.2ex}
                                   {-1em}
                                   {\normalfont\normalsize\bfseries}}

\makeatother

\newcommand{\be}{\begin{equation}}
\newcommand{\ee}{\end{equation}}
\newcommand{\bea}{\begin{eqnarray}}
\newcommand{\eea}{\end{eqnarray}}
\newcommand{\ba}{\begin{array}}
\newcommand{\ea}{\end{array}}

\newcommand{\id}{\hbox{1\kern-.27em l}}
\newcommand{\ZZ}{\mathbb{Z}}

\newcommand{\ad}[1]{\mathrm{Ad}_{#1}}

\newcommand{\al}{\alpha}
\newcommand{\ga}{\gamma}
\newcommand{\Ga}{\Gamma}
\newcommand{\bet}{\beta}

\newcommand{\ka}{\kappa}

\newcommand{\ep}{\epsilon}
\newcommand{\si}{\sigma}
\newcommand{\la}{\lambda}

\newcommand{\om}{\omega}

\newcommand{\La}{\Lambda}

\newcommand{\cN}{\mathcal{N}}

\newcommand{\cM}{\mathcal{M}}
\newcommand{\cA}{\mathcal{A}}
\newcommand{\cL}{\mathcal{L}}
\newcommand{\cD}{\mathcal{D}}

\newcommand{\cP}{\mathcal{P}}

\newcommand{\SU}{\mathrm{SU}}

\newcommand{\SO}{\mathrm{SO}}
\newcommand{\SL}{\mathrm{SL}}
\newcommand{\Sp}{\mathrm{Sp}}
\newcommand{\Spin}{\mathrm{Spin}}

\newcommand{\g}{\mathfrak{g}}
\newcommand{\su}{\mathfrak{su}}
\renewcommand{\u}{\mathfrak{u}}
\newcommand{\so}{\mathfrak{so}}

\renewcommand{\sp}{\mathfrak{sp}}

\newcommand{\iso}{\cong}

\begin{document}

\begin{center}

\vspace*{5mm}
{\Large\sf The Weak Coupling Spectrum around Isolated Vacua in \\$\cN = 4$ Super Yang-Mills on $T^3$ with any Gauge Group}

\vspace*{5mm}
{\large Josef Lindman H\"ornlund, Fredrik Ohlsson}

\vspace*{5mm}
Department of Fundamental Physics\\
Chalmers University of Technology\\
S-412 96 G\"oteborg, Sweden\\[3mm]
{\tt lindman@student.chalmers.se\\fredrik.ohlsson@chalmers.se}     
     
\vspace*{5mm}{\bf Abstract:} 
                                 
\end{center}

\noindent The moduli space of flat connections for maximally supersymmetric Yang-Mills theories, in a space-time of the form $T^3 \times \mathbb{R}$, contains isolated points, corresponding to normalizable zero energy states, for certain simple gauge groups $G$. We consider the low energy effective field theories in the weak coupling limit supported on such isolated points and find that when quantized they consist of an infinite set of harmonic oscillators whose angular frequencies are completely determined by the Lie algebra of $G$. We then proceed to find the isolated flat connections for all simple $G$ and subsequently specify the corresponding effective field theories. 

\pagenumbering{arabic}

\setcounter{equation}{0}
\input{Introduction}

\setcounter{equation}{0}
\input{TopologicalAspectsInYangMillsTheory}

\setcounter{equation}{0}
\input{LowEnergyEffectiveTheories}

\setcounter{equation}{0}
\input{SUn}

\setcounter{equation}{0}
\input{Spinn}

\setcounter{equation}{0}
\input{Spn}

\setcounter{equation}{0}
\input{Exceptional}

\setcounter{equation}{0}
\input{Summary}

\section*{Acknowledgement}
The authors would like to thank M\aa ns Henningson for providing the problem and invaluable guidance and Niclas Wyllard for many helpful and enlightening discussions. Furthermore we are thankful to the JHEP referee for several valuable comments on the material in this paper and on improvements of its presentation. Finally, we would also like to acknowledge Ulf Gran for an introduction to gamma matrix computer algebra. Fredrik Ohlsson is supported by a grant from the G\"oran Gustafsson foundation. Josef Lindman H\"ornlund would like to thank the Department of Fundamental Physics at Chalmers for hospitality during the writing of his master thesis. 




\end{document}

%% file: Introduction.tex
\section{Introduction}
An $\cN = 4$ supersymmetric Yang-Mills theory in (3+1) space-time dimensions is completely defined by the Lagrangian density \cite{Brink:1977}
\bea
\label{eqn:SYMLagrangianDensity}
\cL & = & \mathrm{Tr} \left\{ -\frac{1}{g^2} \left( \frac{1}{4} F_{\mu\nu}F^{\mu\nu} - \frac{i}{2} \overline{\Psi}^I\Ga^{\mu}D_{\mu}\Psi_I - \frac{1}{2} D_{\mu}\Phi_AD^{\mu} \Phi^A \right. \right. \nonumber \\
&& \left. \left. +\frac{1}{4} [\Phi_A,\Phi_B][\Phi^A,\Phi^B] + \frac{i}{2} \overline{\Psi}^I\Ga^A[\Phi_A,\Psi_I] \right) + \frac{\theta}{8\pi^2} \epsilon^{\mu\nu\rho\sigma}F_{\mu\nu}F_{\rho\sigma} \right\} \,,
\eea 
a gauge group $G$ and the values of the theta angle $\theta$ and the coupling constant $g$. The topological $\theta$-term plays no part in the following considerations and will therefore be excluded. The field content of the theory is a gauge field $A^{\mu}$, six scalar fields $\Phi^A$ and four Majorana spinors $\Psi^I$ ($\overline{\Psi}^I$), where $A$ and $I$ are indices in the $\mathbf{6}$ and $\mathbf{4}$ ($\mathbf{\overline{4}}$) representations of $\SO(6)$ respectively. These fields constitute a vector multiplet of the four supersymmetry generators with eight bosonic and eight fermionic degrees of freedom on-shell. We will consider this theory in a space-time of the form $T^3 \times \mathbb{R}$.

The aim of this paper is to study the effective theory around certain zero energy states in the weak coupling limit. In the temporal gauge $A^0 \equiv 0$, which will be used throughout this article, vacuum states are characterized by a gauge field with vanishing magnetic and electric contributions to the energy, proportional to $\mathrm{Tr}(F^{ij}F^{ij})$ and $\mathrm{Tr}(F^{0i}F^{0i})$ respectively. A gauge field $A^i$ with vanishing spatial components $F^{ij}$ of the field strength, i.e. zero curvature, is called a flat connection. Zero energy states are thus supported on the moduli space, $\cM$, of gauge inequivalent flat connections. The momentum conjugate to $A^i$ is $F^{0i}$, implying that a low energy state is locally constant on $\cM$. In general, $\cM$ will be disconnected and in some particular cases it will even contain isolated points.

As we will see in the following section, the moduli space $\cM$ is parameterized by conjugacy classes of triples of elements of $G$, known as almost commuting triples, commuting amongst themselves up to possible multiplication with elements of the center of $G$. Isolated conjugacy classes of commuting triples where first studied in their own right in mathematics \cite{Borel:1953,Borel:1961}. Later their application in physics, through their connection to isolated points in the moduli space of vacua in supersymmetric Yang-Mills theory, was discovered by Witten in \cite{Witten:1998} where the orthogonal groups were treated. Later, this analysis was extended in \cite{Keurentjes:1998} and a complete classification of commuting triples and the corresponding moduli spaces, was provided in \cite{Kac:1999,Keurentjes:1999a,Keurentjes:1999b}. Shortly thereafter a complete classification of almost commuting triples was given by Borel et. al. \cite{Borel:1999}.


In addition to a locally constant flat connection, the vacuum states are characterized by covariantly constant scalar and spinor fields. At each point in $\cM$, corresponding to a zero energy gauge field configuration, there is an additional vector space of vacuum states associated to the scalar and spinor degrees of freedom. For the isolated points in $\cM$ these additional vector spaces are one-dimensional. Hence, fixing an isolated flat connection completely specifies a vacuum state, which will turn out to be normalizable. 

In fact, the focus of this paper will be the field theories localized at isolated vacua. Such theories are obtained by considering the vacuum field configuration as a background and expanding the fields to lowest order in the coupling constant around it. The equations of motion for the fluctuations are found to be completely solvable and the solutions expressible in a basis of eigenfunctions of the covariant derivative, whose spectrum is completely determined by the Lie algebra of $G$. The effective theory is found to be described by an infinite set of quantum mechanical harmonic oscillators, corresponding to the possible excitations. The angular frequencies, or equivalently the excitation energies, of the oscillators are related to the eigenvalues of the covariant derivative. Hence, the Lie algebra of $G$ completely determines the effective theory.

The outline of this paper will be the following: In section two we consider relevant topological aspects of the principal gauge bundles corresponding to the points in $\cM$. In section three we derive the low energy effective theory at the isolated points in $\cM$ and its dependence on the Lie algebra of $G$. In sections four through seven we then apply the method developed in sections two and three to specify the low energy theory for all simple gauge groups $G$ containing isolated vacua.

%% file: TopologicalAspectsInYangMillsTheory.tex
\section{Topological Aspects in Yang-Mills Theory}
Due to the choice of temporal gauge, the fields of the $\cN=4$ theory can be treated as sections of various bundles over $T^3$. From this point of view the gauge field $A^i$ is the connection of a principal $G_{adj}$-bundle, where $G_{adj} = G / C_G$ is the adjoint form of a connected, simply connected Lie group $G$. Here, and for the remainder of this paper, we use $C_G$ to denote the center of a group $G$. The scalar and spinor fields are then sections of associated $ad$-bundles and topological considerations are therefore restricted to the principal $G_{adj}$-bundle. 

\subsection{Topological configurations}
\label{sec:Topology}
\begin{figure}[!ht]
\begin{center}
\begin{overpic}[scale=0.8]{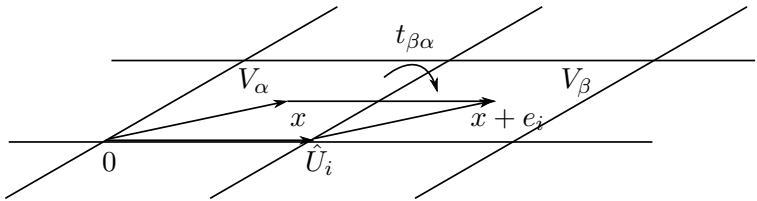}
\put(31,15){$V_{\alpha}$}
\put(74,15){$V_{\beta}$}
\put(52,21){$t_{\bet\al}$}
\put(38,10){$x$}
\put(62,10){$x + e_i$}
\put(40,4){$\hat{U}_i$}
\put(13,4){$0$}
\end{overpic}
\caption{Illustration of coordinate patches and associated transition functions for the two-torus. Two homotopically equivalent curves from the origin to $x+e_i$ are also shown.}
\label{fig:TorusPatches}
\end{center}
\end{figure}

The base manifold $B$ of a bundle must generally be described using a set of open coverings $\{V_{\al}\}$. The topology of the bundle is then specified through transition functions, $t:B \rightarrow G_{adj}$, on the sections of overlapping such patches. The three-torus may be defined as $\mathbb{R}^3/\La$, where $\La$ is the span over $\ZZ$ of three lattice vectors $e_{i=1,2,3}$. We assume, without loss of generality, that the torus is $\mathbb{R}^3/\mathbb{Z}^3$, i.e. that the lattice vectors $e_i$ are orthonormal. We will treat $\mathbb{R}^3$ as the base manifold and impose periodicity conditions on sections from $\mathbb{R}^3$ to the corresponding fibers. In this way we effectively obtain fields over a three-torus for which the transition functions appear as twisted periodicity conditions of the form
\be
\left\{ \ba{ccl}
A_{\al} & = & t_{\bet\al}^{-1} A_{\bet} t_{\bet\al} + t_{\bet\al}^{-1} \mathrm{d}t_{\bet\al} \\
\phi_{\al} & = & t_{\bet\al}^{-1} \phi_{\bet} t_{\bet\al}
\ea \right.
\ee
in the principal and adjoint bundles respectively. Every patch in $\{V_{\al}\}$ is thence a cell in the lattice and the overlap between adjacent patches are the two-faces $F_{i=1,2,3}$ of the lattice cells. We will use the convention, as illustrated for a two-torus in Figure \ref{fig:TorusPatches}, that translation with a lattice vector $e_i$ corresponds to going from $V_{\al}$ to $V_{\bet}$ with the transition function $t_{\bet\al}$, which then depends only on the coordinates on $F_i$. Subsequently, we will simply denote the transition functions corresponding to the $e_i$ translation by $t_i:F_i \rightarrow G_{adj}$.

Consider now the general case with non-trivial transition functions on all three two-faces $F_i$ of the torus. Due to the triviality of $\pi_2(G_{adj})$ and $\pi_0(G_{adj})$ for all connected Lie groups, the transition functions are completely characterized by the homotopy classes, which are elements of $\pi_1(G_{adj}) \iso C_G$, of the three homotopically inequivalent closed non-contractible curves $\ga_{i=1,2,3}$ on $T^3$ with a common base point. We note that $\ga_i$ are the generators of $\pi_1(T^3)$. These homotopy classes form a triplet $\hat{m}' \in {C_G}^3$ encoding the bundle topology.

An equivalent way to describe the topological class of a bundle is to introduce the concept of a {\it holonomy} 
\be
\label{eqn:gTilde}
\tilde{g}(x) \equiv \cP \left( \exp \int_{0}^x A_i dx^i \right) \,,
\ee
along a curve in $\mathbb{R}^3$, where $\cP$ denotes path-ordering. Under bundle automorphisms, the holonomy $\tilde{g}(x)$ transforms by conjugation. Fixing a flat connection $A^i$ and calculating holonomies around the curves $\ga_i \ga_j \ga^{-1}_i \ga^{-1}_j$ then yields a triplet $(\id,\id,\id) \in {G_{adj}}^3$ since these curves are all contractible. When lifted to $G$, however, this triplet is mapped to a triplet $\hat{m} \in {C_G}^3$. This procedure therefore induces a homomorphism from the isomorphism classes of $G_{adj}$-bundles over $T^3$ to ${C_G}^3$, just as $\hat{m}'$ did. The triplet $\hat{m}$ transforms as a vector under the mapping class group $\SL(3,\ZZ)$ of $T^3$ \cite{Henningson:2008}. For a cyclic center $C_G$ it is possible to put $\hat{m}$ on the form $(\id,\id,m)$ using the $\SL(3,\ZZ)$-transformations, and when able we will only use the third $\hat{m}$-component to characterize the non-triviality of the bundle. Generically, only one of the components can be put to the identity. In the following this is, however, only the case for $G=\Spin(4n)$ where we will retain the full $\hat{m}$-vector when necessary.

\subsection{Almost commuting triples}
We now return to the moduli space $\cM$ of flat connections, introduced in the previous section, which can be parametrized by the holonomies 
\be
\label{eqn:Ui}
\hat{U_i} = \cP \left( \exp \int_{\ga_i} A_i dx^i \right) \,,
\ee
around the three homotopically different one-cycles $\ga_i$. Hence, every flat connection corresponds to a triple $(\hat{U}_1,\hat{U}_2,\hat{U}_3)$ of commuting elements in $G_{adj}$. Using these holonomies $\hat{U}_i$ we can express $\tilde{g}$ along a generic curve, see Figure \ref{fig:TorusPatches}, from the origin to a point $x^i+e_j^i$ on an adjacent patch as
\be
\label{eqn:PeriodicityGTilde}
\tilde{g}(x^i + e^i_j) = \hat{U}_j \tilde{g}(x^i) t_j(x^i) \,.
\ee 
This expression will prove useful in later considerations.

The unique lift of (\ref{eqn:Ui}) to the covering group $G$ defines a triple $(U_1,U_2,U_3)$ of {\it almost commuting} elements, i.e. elements satisfying
\be
\label{eqn:LiftedUi}
m_{ij} = U_i U_j U_i^{-1} U_j^{-1} \,,
\ee  
where $m_{ij} \in C_G$ are the elements of a triple $\hat{m} = (m_{23},m_{31},m_{12})$ previously described. The $m_{ij} \neq \id$ are the obstructions to lift the flat $G_{adj}$-bundle to a $G$-bundle. The bundle topology is thus described by the commutation relations (\ref{eqn:LiftedUi}) among the lifted holonomies $\{U_i\}$. In order to constitute well defined coordinates on $\cM$ the triples (\ref{eqn:Ui}) in $G_{adj}$ must reflect the equivalence of flat connections that are related through gauge transformations. Therefore, we must consider triples related through simultaneous conjugation by an arbitrary element $g\in G_{adj}$ as equivalent. The coordinates on $\cM$ are thus the conjugacy classes of commuting triples $[\{\hat{U}_i\}]$ in $G_{adj}$. In fact, due to the uniqueness of the lift of $\hat{U}_i$ it is possible to use the conjugacy classes $[\{U_i\}]$ of the lifted holonomies to parametrize $\cM$. 

When considering the moduli space $\cM$ for a certain gauge group we will work with the lifted holonomies in $G$ and index the moduli space with $G$ rather than $G_{adj}$. On the other hand, when considering the field theory localized at a certain points in the moduli space we will instead use the holonomies $\hat{U}_i$ to emphasize the fact that $G_{adj}$ is the gauge group of the original supersymmetric Yang-Mills theory. When it is clear from the context that the lifted holonomies are considered we will refer to triples $\{U_i\}$ with $m_{ij}=\id$ as commuting.

We can now consider all triples obeying (\ref{eqn:LiftedUi}) when lifted to $G$ for a given $\SL(3,\ZZ)$ equivalence class $[\hat{m}]$ of vectors in ${C_G}^3$ as describing a subspace $\cM([\hat{m}]) \subset \cM$. These disjoint subspaces are generally themselves disconnected,
\bea
\label{eqn:MmComponents}
\cM([\hat{m}]) = \underset{a}{\bigcup} \, \cM_{a} &,& \hat{m} \in {C_G}^3 \,.
\eea

A triple may break generators of $\mathrm{Lie}\ G$ and the unbroken gauge group $H \subset G_{adj}$ is defined as the commutant of the triple $\{\hat{U}_i\}$, i.e. the subgroup of $G_{adj}$ corresponding to the unbroken generators. On each component $\cM_{a}$ of $\cM([\hat{m}])$, the rank $r_a$ of $H$ is constant and for trivial topology $\mathrm{rank}(H) = \mathrm{rank}(G_{adj})$ on the identity component of $\cM([\hat{m}])$. There is a remarkable relation between the ranks $r_a$ and the dual Coxeter number $g^{\vee}$ of $G$,
\be
\label{eqn:CoxeterIdentity}
{\displaystyle \sum_a} (r_a + 1) = g^{\vee}\,,
\ee
conjectured in \cite{Witten:1998} and proven in theorems 1.4.1 and 1.5.1 of \cite{Borel:1999}.

At a generic point in $\cM([\hat{m}])$, $H$ consists only of abelian factors, i.e. $\mathrm{Lie}\ H = \mathfrak{h} = \u(1)^{r_a}$. However, at certain subspaces $\cM^H$ of $\cM([\hat{m}])$ this symmetry may be enhanced with non-abelian factors to $\mathfrak{h} = \mathfrak{s} \varoplus \u(1)^r$, where $r < r_a$ and $\mathfrak{s}$ is semi-simple. In the following section we will find that triples with rank zero commutant, i.e. belonging to subspaces of $\cM$ where $H$ is finite, are of special interest. We will now proceed to show that all such points are isolated in $\cM$ and vice versa that all isolated triples break the gauge group completely. To do so we first introduce a convenient basis $T_{\la}$ of $\mathrm{Lie}\ G$ satisfying
\be
\label{eqn:TLambda}
U_i^{-1} T_{\la} U_i = \la_i T_{\la} \,,
\ee
where the components $\la_i$ form a vector $\vec{\la}$ of eigenvalues. Observe that it is always possible to find such a basis since the holonomies are mutually almost commuting. The action of the mapping class group $\SL(3,\ZZ)$ on the holonomies induces an action on the eigenvalue vectors $\vec{\la}$. For trivial topology, where the triples are on equal footing, the spectrum of eigenvalue vectors will be $\SL(3,\ZZ)$-invariant. For non-trivial topology this invariance is broken, e.g. for a triple on the standard form it is reduced to an $\SL(2,\ZZ)$-invariance in the untwisted directions. We also note that a rank zero triple has no eigenvalue vector $\vec{\la} = (1,1,1)$ since all generators of the Lie algebra are broken.

The proof that the properties having rank zero and being isolated in $\cM$ are equivalent then goes as follows: First consider an infinitesimal perturbation of an almost commuting triple
\be
\label{eqn:CommutingtriplePerturbation}
U'_i = U_i(\id + \ep_i^{\la}T_{\la})\,.
\ee
The requirement that the commutation relations among the $U'_i$ are the same as among the $U_i$ implies the condition
\be 
\label{eqn:CommutationCondition}
\ep_i^{\la} (\la_j - 1) = \ep_j^{\la} (\la_i - 1)
\ee
on the $\ep_i$. Under the assumption that $\vec{\la} \neq (1,1,1)$ there is always one component $\la_i \neq 1$, which we can choose to be $\la_3$ by relabeling the $U_i$'s if necessary. Using the condition (\ref{eqn:CommutationCondition}) the original perturbation can then be written as
\be
\label{eqn:GlobalTransformation}
U'_i = U_i + \frac{\ep^{\la}_3}{\la_3 - 1} [T_{\la},U_i] \,,
\ee
which constitutes an infinitesimal gauge transformation, implying that $U'_i$ is in the same conjugacy class as $U_i$. Thus, $\{U_i\}$ cannot be perturbed to an inequivalent triple and is hence isolated in the moduli space $\cM$. The converse is proved by assuming that there is a $\vec{\la} = (1,1,1)$ eigenvalue vector. Repeating the argument above then leads to a contradiction.

\subsubsection{Construction of the moduli space}
\label{sec:ModuliSpaceConstruction}
From the results in \cite{Borel:1999} we have a recipe for finding the structure of $\cM([\hat{m}])$ in (\ref{eqn:MmComponents}), for any gauge group and any $\hat{m}$-vector. This method involves studying the extended Dynkin diagram $\widetilde{D}$ in the case of trivial $\hat{m}$ and its quotient counterpart for $\hat{m}$ non-trivial. The quotient diagram $\widetilde{D}/\si$ is constructed from the extended Dynkin diagram by identifying all nodes in each orbit under a diagram automorphism $\si:\widetilde{D} \rightarrow \widetilde{D}$. The automorphisms form a group $\Sigma$ which is isomorphic to the center $C_G$. From this it can be argued that there is a correspondence between the automorphism $\si$ and the $\SL(3,\ZZ)$ equivalence class $[\hat{m}]$ encoding the bundle topology. We denote by $g_{\al}$ the coroot integers of $\widetilde{D}$, also known as the dual Coxeter labels, and by $g_{\overline{\al}}$ the quotient coroot integers of $\widetilde{D}/\si$. The $g_{\overline{\al}}$ are simply the sum of the coroot integers of the nodes in the $\si$-orbit or equivalently the original coroot integers multiplied by the cardinality of their respective orbits.

To find every disconnected component of $\cM([\hat{m}])$ we consider all integers $k \in [1,\max_{\scriptscriptstyle \widetilde{D}}(g_{\al})]$ dividing at least one $g_{\al}$ of $\widetilde{D}$ or $k \in [1,\max_{\scriptscriptstyle \widetilde{D}/\si}(g_{\overline{\al}})]$ dividing at least one $g_{\overline{\al}}$ of $\widetilde{D}/\si$ for $\hat{m}$ trivial or non-trivial respectively. Each $k$ corresponds to $\varphi(k)$ components $\cM_a$ with rank $r_a$, where $r_a+1$ is equal to the number of $g_{\al}$ or $g_{\overline{\al}}$ divisible by $k$. Here $\varphi(k)$ is the Euler $\varphi$-function, given by the number of integers less than or equal to $k$ that are coprime to $k$. Furthermore, we call $k$ the {\it order} of the triples of the component $\cM_a$. For commuting rank zero triples the order $k$ is simply the order of the elements $U_i$ in $G$. For almost commuting rank zero triples this is not the case. Instead, for groups with cyclic center, $k$ refers to the order of the third element $U_3$ when viewed as a component in the centralizer $\mathcal{Z}$ of $U_1$ and $U_2$. This perspective is possible since $\pi_0(\mathcal{Z})$ has a cyclic structure, and the components of $\mathcal{Z}$ are in one to one correspondence with the conjugacy classes by the Lemma 9.1.2 in \cite{Borel:1999}. Again, groups with non-cyclic centers are exceptions. The moduli space components obtained using this method obey (\ref{eqn:CoxeterIdentity}) by construction and in sections four through seven it will be used to construct the moduli spaces $\cM_G$ of all simple Lie groups $G$. 

\begin{figure}[!ht]
\begin{center}
\includegraphics[scale=0.3]{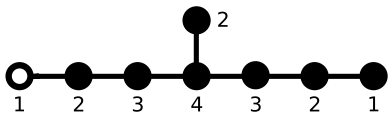}
\caption{Extended Dynkin diagram of $E_7$}
\label{fig:DynkinE7Extended}
\end{center}
\end{figure}

As an illustrative example we apply the method of moduli space construction to the case $G = E_7$. The center $C_{E_7}$ is isomorphic to $\ZZ_2$ which corresponds to an extended Dynkin diagram, shown in Figure \ref{fig:DynkinE7Extended}, that is invariant under a finite set of diagram automorphisms $\Sigma$ consisting of the identity mapping and reflection in the node with $g_{\al} = 4$. As $C_{E_7}$ is cyclic we put $\hat{m}$ on the standard form $(\id,\id,m)$ and specify the topology by $m=\pm \id$.

We begin with the trivial $m=\id$ case. Studying the extended Dynkin diagram in Figure \ref{fig:DynkinE7Extended} we find that $k=1$ divides all eight $g_{\al}$ and thus results in the maximal rank identity component. Furthermore, $k=2$ divides four labels, $k=3$ divides two and $k=4$ divides a single label. The Euler $\varphi$-function takes the values $1,1,2$ and $2$ for the four $k$-values respectively. The moduli space for this case is thus
\be 
\label{eqn:ModuliSpaceE7Identity}
\cM_{E_7}(m=\id) = \cM_7 \cup \cM_3 \cup \cM_1 \cup \cM '_1 \cup \cM_0 \cup \cM '_0 \,,
\ee
where $\cM_{r_a}$ denotes a component of rank $r_a$. 

\begin{figure}[!ht]
\begin{center}
\includegraphics[scale=0.3]{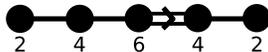}
\caption{Quotient diagram of $E_7$}
\label{fig:DynkinE7Quotient}
\end{center}
\end{figure}

The quotient diagram is formed by dividing out the only non-trivial automorphism, producing the diagram in Figure \ref{fig:DynkinE7Quotient}. 

Clearly $k=1,2$ divide all five labels $g_{\overline{\al}}$, $k=3,6$ divide one and $k=4$ divides two. For these $k$, $\varphi(k)$ takes the values $1,1,2,2$ and $2$ respectively. The contributions from all $k$ make up the complete moduli space
\be
\label{eqn:ModuliSpaceE7NegativeIdentity}
\cM_{E_7}(m=-\id) = \cM_4 \cup \cM '_4 \cup \cM _1 \cup \cM '_1 \, \overset{4}{\underset{i=1}{\bigcup}} \, \cM^{(i)}_0 \,.
\ee

\subsubsection{Conjugacy classes and the action of the center}
\label{sec:ConjugacyClasses}
We will now make a detour, that will prove useful later on, to study the problem of determining, for arbitrary gauge group $G$, how the conjugacy classes of rank zero triples, corresponding to the isolated points of $\cM([\hat{m}])$, are related and how the center $C_G$ acts on the set of conjugacy classes. These problems are treated in full detail in \cite{Borel:1999}. We will, however, restrict our expos\'e to stating results that will prove useful for our purposes.  

We start with considering the case of commuting rank zero triples of order $k$ in $G$. According to Proposition 5.1.5 in \cite{Borel:1999} there is a relation between the $\varphi(k)$ conjugacy classes of such triples. If $(U_1,U_2,U_3) \in G$ is a representative of a conjugacy class, then representatives of all other conjugacy classes are obtained as $(U_1,U_2,U_3^l)$, where $l$ is coprime to $k$. We define the action of an element $\ga \in C_G$, with respect to e.g. the third component of the triple as $\ga \cdot (U_1,U_2,U_3) = (U_1,U_2,\ga U_3)$. The induced action on the set of conjugacy classes is trivial, i.e. is the identity mapping for all $\ga$. 

The case of almost commuting rank zero triples of order $k$ is somewhat more involved, since constructing representatives of all the $\varphi(k)$ classes is no longer straightforward using any one representative. However, the action of the center on the set of conjugacy classes can be used to deduce some information about the relation between them. In the following we will assume that $\hat{m}$ is on standard form, i.e. that the center $C_G$ is cyclic. According to Lemma 9.1.10 in \cite{Borel:1999} the induced action of the center on the set of conjugacy classes is trivial with respect to $U_1$ and $U_2$. However, according to Lemma 9.1.12 the stabilizer $\mathcal{K}$ of the induced action is not necessarily all of $C_G$ when $\ga$ is acting on the third component. The action of $\ga \notin \mathcal{K}$ thus relates representatives of distinct conjugacy classes. Finding $\mathcal{K}$ is in general a complicated process and therefore we will simply state it in the cases where we need to invoke the action of the center in our arguments.

To summarize; for all groups we can always choose representatives of the conjugacy classes of almost commuting rank zero triples so that they have the same $U_1$ and $U_2$ components. Hence the third component $U_3$ completely determines the conjugacy class.

%% file: LowEnergyEffectiveTheories.tex
\section{Low energy effective theories}
\label{sec:LowEnergyEffectiveTheories}
We will now consider the field theories localized at subspaces $\cM^H \subset \cM$, with unbroken gauge group $H$. The theory at $\cM^H$ is completely characterized by $H$. However, when $H$ is not semi-simple the vacuum states of $\mathcal{N} = 4$ Yang-Mills theory are not normalizable, since the abelian degrees of freedom correspond to free fields, or equivalently, plane waves. Finite theories are therefore localized at $\cM^H$ where the unbroken gauge group is semi-simple or at isolated points where $H$ is a finite group. We note that all scalar and spinor field modes corresponding to broken generators acquire energy, since covariantly constant modes can only originate from unbroken generators. They are therefore irrelevant when considering low energies.

At $\cM^H$ with $H$ semi-simple the low energy effective theory can be described by supersymmetric quantum mechanics with gauge group $H$ \cite{Witten:2000}. This theory is conjectured to have a finite number, depending on $H$, of normalizable states which are bound at threshold. However, these states are notoriously elusive and have not been explicitly constructed or even rigorously proven to exist.

\subsection{Weak coupling expansion around vacua}
In the weak coupling limit it is possible to expand the fields around any normalizable zero energy field configuration $\{\cA^{\mu},\phi_0^A,\psi_0^I\}$. To lowest order in the coupling constant $g$ this expansion is
\be
\label{eqn:PerturbativeExpansionOfFields}
\left\{
\ba{ccc}
A^{\mu} & = & \cA^{\mu} + ga^{\mu}\\
\Phi^A & =  & \phi_0^A + g\phi^A\\
\Psi^I & = & \psi_0^I + g\psi^I
\ea
\right. \,.
\ee
Control of the theory resulting from this expansion requires a thorough understanding of the vacuum state $\{\cA^{\mu},\phi_0^A,\psi_0^I\}$. Such an understanding is not yet obtained for the supersymmetric quantum mechanical bound states at threshold. The validity of an expansion around such states is therefore somewhat uncertain, and further investigations are required in order to complete a satisfying argument justifying (\ref{eqn:PerturbativeExpansionOfFields}). In order to avoid complications arising from these states we will therefore restrict our considerations to gauge field configurations where the gauge group has been completely broken, i.e. where $\mathrm{rank}(H) = 0$. In these cases the corresponding quantum mechanics is trivial and adds no complication to the analysis. Furthermore, the $\phi_0^A$ and $\psi_0^I$ modes all vanish in the effective theory, i.e. $\phi_0^A = \psi_0^I = 0$, since no unbroken generators remain. Hence, we proceed to study the expansion (\ref{eqn:PerturbativeExpansionOfFields}) around the isolated points, considered in the previous section, of the moduli space $\cM$ only. We note that the perturbation of the gauge field does not influence the topology of the principal bundle, the connection of the bundle is still $\cA^{\mu}$.

Inserting (\ref{eqn:PerturbativeExpansionOfFields}) in the Lagrangian density (\ref{eqn:SYMLagrangianDensity}) and using the vanishing spatial curvature of $\cA^{\mu}$ we obtain the effective low energy Lagrangian density
\be
\label{eqn:SYMLagrangianDensityExpansion}
\cL = -\frac{1}{2} \mathrm{Tr} \left\{(\cD_{\mu}a_{\nu}\cD^{\mu}a^{\nu} - \cD_{\mu}a_{\nu}\cD^{\nu}a^{\mu})- \cD_{\mu}\phi^A\cD^{\mu}\phi_A - i\psi^I\Ga^{\mu}\cD_{\mu}\psi_I \right\} + \mathcal{O}(g) \,,
\ee
where $\cD^{\mu}$ is the covariant derivative with respect to $\cA^{\mu}$. Defining the differential operators $\Delta_0 \equiv \cD_i \cD^i$ and $\Delta_{1/2} \equiv \Ga^0 \Ga^i \cD_i $ the equations of motion for the scalar and spinor fields in this minimally coupled expression are
\bea
\label{eqn:EQMphi}
\ddot{\phi}^A + \Delta_0 \phi^A & = & 0 \\
\label{eqn:EQMpsi}
\dot{\psi}^I + \Delta_{1/2} \psi^I & = & 0 \,.
\eea
Note that the equations (\ref{eqn:EQMpsi}) do not mix the chiral and anti-chiral parts of $\psi^I$ due to the appearance of $\Ga^0$ in $\Delta_{1/2}$.

For the fluctuation $a^{\mu}$ in the gauge field there is a redundancy in the space of solutions to the equations of motion, since all fields of the form $a^{\mu} = \cD^{\mu} \la$, where $\la$ is an arbitrary scalar field, are trivially on shell. This redundancy constitutes a gauge invariance under a transformation $a^{\mu} \mapsto a^{\mu} + \cD^{\mu} \la$, since the connection $\cA^{\mu}$ is flat. Because of this invariance, we must restrict ourselves to divergence free fluctuations, i.e. $a^{\mu }$ satisfying $\cD_j a^j=0$, in order to only consider gauge inequivalent solutions. Such fields form a subspace of Lie algebra-valued 1-forms on which the equations of motion are restricted to
\be
\label{eqn:EQMa}
\ddot{a}^j + \Delta_0 a^j = 0 \,.
\ee
We note that the fluctuation $a^{\mu}$ and the scalar fields $\phi^A$ obey the same equations of motion. The choice of gauge $a^0=0$ and the requirement $\cD_i a^i=0$ reduce the number of independent $a^{\mu}$ components to two, preserving the equal numbers of fermionic and bosonic degrees of freedom after the expansion (\ref{eqn:PerturbativeExpansionOfFields}).

The differential operator $i\cD_i$, appearing in the equations of motion, is self-adjoint with spectrum
\be
\label{eqn:DiSpectrum}
i \cD_i \varphi_m(x^j) = (\om_m)_i \varphi_m (x^j) \,,
\ee
when acting on scalar functions, where the eigenvalues $(\om_m)_i$ are guaranteed to be real. Since the torus is a compact manifold, the eigenfunctions $\varphi_m$ span the space of Lie algebra-valued functions on $T^3$. This implies that the spatial dependence of all fields in the expanded theory can be expressed in terms of the eigenfunctions $\varphi_m$ through the expansions
\bea
\label{eqn:FourierExpansionA}
a^j(x) & = & {\displaystyle \sum_m a^j_m(t) \varphi_m(x^i)}\\
\label{eqn:FourierExpansionPhi}
\phi^A(x) & = & {\displaystyle \sum_m \phi^A_m(t) \varphi_m(x^i)}\\
\label{eqn:FourierExpansionPsi}
\psi^I(x) & = & {\displaystyle \sum_m \psi^I_m(t) \varphi_m(x^i)} \,,
\eea
where $a^j_m(t)$ are vectors in the two-dimensional space of gauge fields, $\phi^A_m(t)$ are vectors in the six-dimensional space of scalar fields and $\psi^I_m(t)$ are vectors in the four-dimensional space of Majorana spinors. Inserting the expansions into the equations of motion and using the completeness of the set $\{\varphi_m\}$ to integrate out the spatial dependence we can solve for $a^j_m(t)$, $\phi^A_m(t)$ and $\psi^I_m(t)$. The solutions, when quantized, correspond to one set of creation and annihilation operators of a harmonic oscillator for each degree of freedom. All the oscillators have the same angular frequency $\om_m$, related to the eigenvalues of $i\cD_i$ through 
\be
\label{eqn:AngularFrequency}
\om_m \equiv \left({\displaystyle \sum_{i=1}^3 \, (\om_m)_i^2}\right)^{1/2} \,.
\ee
Thus, the Hamiltonian
\be 
\label{eqn:ExpandedTheoryHamiltonian}
H = {\displaystyle \sum_m} \left\{ \hbar \om_m{\displaystyle \sum_{i=1}^8 \left( (N_b)^i_m + \frac{1}{2} \right)} + \hbar \om_m{\displaystyle \sum_{i=1}^8 \left( (N_f)_m^i - \frac{1}{2} \right)} \right\}
\ee
of the effective theory is that of eight bosonic and eight fermionic harmonic oscillators, of equal angular frequency $\om_m$, for each eigenfunction $\varphi_m(x^i)$ of $i\cD_i$. Here, $N_b$ and $N_f$ are the number operators, counting the level of excitations of the bosonic and fermionic oscillators respectively. From the Hamiltonian it is immediately clear that the ground state of the theory, i.e. the state with $(N_b)^i_m = (N_f)^i_m = 0$, has zero energy as the energies of the oscillators cancel for each $m$. Note that the general approach to expand the fields of the theory in eigenfunctions of the $i\cD_i$ operator is applicable in all supersymmetric Yang-Mills theories on $T^3$. The general form of the Hamiltonian will be the same, with the index $i$ in (\ref{eqn:ExpandedTheoryHamiltonian}) e.g. taking the values $1 , \ldots , 4$ in the $\cN = 2$ case.

From the Hamiltonian (\ref{eqn:ExpandedTheoryHamiltonian}) we find that the structure, i.e. the available states, of the effective field theory at the isolated vacuum state specified by $\cA^{\mu}$ is completely determined by the spectrum of the $i\cD_i$ operator. Hence, in order to characterize the space of states, we proceed with an analysis of this operator and its eigenfunctions. 

\subsection{Construction of eigenfunctions}
In this section we present a general method for constructing the eigenfunctions $\varphi_m(x^i)$ in (\ref{eqn:DiSpectrum}) for any isolated vacuum in $\cM_G$, with $G$ any gauge group. We can construct continuous such solutions using parallel transport of the Lie algebra basis elements $T_{\la}$ of (\ref{eqn:TLambda}). We note that the basis elements have the same spectrum under the adjoint actions of $\{U_i\}$ and $\{\hat{U_i}\}$ since the lifting procedure has no influence on the eigenvalues $\la_i$.

Multiplying the parallel transported field by an exponential function gives a solution to (\ref{eqn:DiSpectrum}) of the form
\be
\label{eqn:phiLambda} 
\phi_{\la}(x^i) = \tilde{g}^{-1}(x^i) T_{\la} \tilde{g}(x^i) e^{\mathrm{log} \la_i x^i} \,, 
\ee
where $\tilde{g}(x^i)$ is as in equation (\ref{eqn:gTilde}). Below, we will argue that any scalar function on $T^3$, periodic up to a gauge transformation, can be expressed in terms of these $\phi_{\la}$. Before proceeding we verify that $\phi_{\la}$ satisfies the periodicity conditions 
\be
\label{eqn:phiPeriodicity}
\phi_{\la}(x^i + e_j^i) = t^{-1}_j(x^i) \phi_{\la}(x^i) t_j(x^i) \,,
\ee
due to property (\ref{eqn:PeriodicityGTilde}), and is indeed an eigenfunction of the operator $i\cD_i$ with
\be
i\cD_i \phi_{\la} = - i \mathrm{log} (\la_i) \phi_{\la} \,.
\ee
Since the holonomies $\{\hat{U}_i\}$ are all group elements with a well defined order, the eigenvalues of $\ad{\hat{U}_i}$ in (\ref{eqn:TLambda}) are all complex roots of unity, i.e. with absolute value one. The logarithm, $\mathrm{log} \la_i$, will therefore be the infinite set of imaginary numbers $i(\mathrm{Arg} \la_i + 2\pi k_i)$, where $k_i \in \ZZ$ and $\mathrm{Arg} \la_i$ is the principal argument of $\la_i$. This means that the set of functions $\{\phi_{\la}\}$ consists of eigenfunctions $\phi_m$ of $i \cD_i$ with real eigenvalues $(\om_m)_i = (\mathrm{Arg} \la_i + 2\pi k_i)$, using the collective index $m = \{\la,k_i\}$ to indicate both the generator $T_{\la}$ and the branch of $\mathrm{log}(\la_i)$.

The argument that (\ref{eqn:phiLambda}) span all Lie algebra-valued fields satisfying the twisted boundary conditions (\ref{eqn:phiPeriodicity}) goes as follows: Let $\xi(x^i)$ be a such a scalar field. We can write $\xi(x^i) = \tilde{g}^{-1}(x^i) T(x^i) \tilde{g}(x^i)$, where $T$ is a Lie algebra-valued function determined uniquely by $\xi$. The periodicity conditions on $\xi$ impose, through the property (\ref{eqn:PeriodicityGTilde}), the condition that 
\be
\label{eqn:PeriodicityConditionT}
\hat{U}_i^{-1} T(x^j+e_i^j) \hat{U}_i = T(x^j)\,.
\ee
We can express $T$ in terms of the $\{T_{\la}\}$ basis as $T(x^i) = \sum_{\la} T_{\la} e^{i \mathrm{Arg} \la_i x^i} f_{\la}(x^i)$ where $f_{\la}$ are scalar functions. The condition (\ref{eqn:PeriodicityConditionT}) then implies that $f_{\la}$ must be a periodic function, $f_{\la}(x^j+e_i^j) = f_{\la}(x^j)$. Such a function can always be written as $f_{\la}(x^i) = \sum_p c_{p,\la} e^{i p_i x^i}$ where $p_i$ is in the reciprocal lattice of $\La$, i.e. $p_i = 2\pi(n_1,n_2,n_3)$ where $n_i \in \ZZ$. Inserting the resulting expression for $T$ into the original expression for $\xi$ and re-introducing the collective index $m = \{\la, n_i\}$ we obtain
\bea
\label{eqn:ArbitraryFieldXi}
\xi(x^i) & = & {\displaystyle \sum_{m}} c_{m} \tilde{g}^{-1}(x^i) T_{\la} \tilde{g}(x^i) e^{i \mathrm{Arg} \la_i x^i + 2 \pi i n_i x^i} \nonumber \\ 
& = & {\displaystyle \sum_{\la} \sum_{k}} c_{\la,k} \tilde{g}^{-1}(x^i) T_{\la} \tilde{g}(x^i) e^{\log^{(k)} (\la_i) x^i}
\,,
\eea
where $(k)$ denotes the branch of $\log{\la_i}$. Hence we see that $\xi$ can be expressed in terms of a linear combination of the $\phi_{\la}$ in (\ref{eqn:phiLambda}), and conclude that $\phi_{\la}$ form a complete set. Thus, the remainder of this article will be concerned with finding the eigenvalues $\la_i$ of (\ref{eqn:TLambda}) in the cases where the gauge group possesses a rank zero triple. 

To summarize, we have found that on the isolated points in the moduli space $\cM$ of flat connections, corresponding to a completely broken gauge group $H$, the low energy effective field theory is an infinite set of quantum mechanical harmonic oscillators with angular frequencies $\om_m$. The isolated flat connections are in one-to-one correspondence with rank zero triples $\{\hat{U}_i\}$. Through the relation
\be
\label{eqn:DiEigenvalues}
(\om_m)_i = (\mathrm{Arg} \la_i + 2\pi k_i) \,,
\ee 
with $k_i \in \ZZ$, the spectrum of the effective theory (\ref{eqn:SYMLagrangianDensityExpansion}) is completely specified by the eigenvalues $\la_i$ of (\ref{eqn:TLambda}) for these triples.

The final part of this paper will be concerned with computing all eigenvalues $\la_i$ for all rank zero triples in all the simple gauge groups $G_{adj}$. We will always work in the universal covering group $G$ of $G_{adj}$ when determining the eigenvalues of the basis $\{T_{\la}\}$ satisfying (\ref{eqn:TLambda}). The procedure, applied for each $G$, will then be to construct the moduli space $\cM_G$, find the almost commuting triples $\{U_i\}$ corresponding to the isolated points and then diagonalize the adjoint action of the $U_i$ on a basis of $\mathrm{Lie}\ G$. According to the remark following (\ref{eqn:ExpandedTheoryHamiltonian}), the results will also be applicable to less-than-maximally supersymmetric Yang-Mills theories. The computed spectra are presented when obtained and summarized in the final section of this paper.

%% file: SUn.tex
\section{The $\SU(n)$ Groups}
\label{sec:Su}
We start applying the method described above to the case $G = \SU(n)$. This will be an important case, since in much of what follows it will be possible to reduce the problem of determining the spectra of various other groups to considerations of  $\SU(n)$ subgroups. Therefore, we will in this section be slightly more explicit in constructing the holonomies and computing the corresponding spectrum. Also, due to certain properties of the special unitary groups we will modify the general method slightly and delay the construction of the moduli space to the very end of this section. 
\begin{figure}[!ht]
\begin{center}
\includegraphics[scale=0.3]{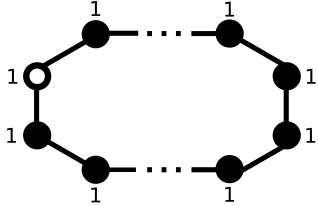}
\caption{Extended Dynkin diagram of $\su(n)$}
\label{fig:DynkinSu}
\end{center}
\end{figure}

From Dynkin diagram considerations we can immediately reject the possibility of finding commuting rank zero triples, i.e. triples with $m = \id$, in $\SU(n)$. This result can also be derived by noting that the semi-simple part of the centralizer of $k$ commuting elements in $\SU(n)$ is a product of special unitary groups, and the orbit of any element under conjugation intersects the maximal torus. Hence, all three elements in the triple can be simultaneously conjugated to the maximal torus, leaving the Cartan generators unbroken.

For each $m \neq \id$ in $C_{\SU(n)} \iso \{z \cdot \id  \mid z^n=1\}$ there exist a pair $(U_1,U_2)$ satisfying the relation
\be
\label{eqn:AlmostCommutingPair}
U_1 U_2 = m U_2 U_1 \,.
\ee
It can be conjugated to the form
\be
\label{eqn:AlmostCommutingPairRepresentation}
\left\{ \ba{ccl}
U_1 & = & \left( \ba{cc}0 & \id_{n-1}\\(-1)^{n-1} & 0\ea \right) \\ 
U_2 & = & a_n \cdot \mbox{diag}(1,z,\dots,z^{n-1})
\ea \right. \,,
\ee
where
\be
a_n = \left\{ 
\ba{ccl}
1 & , & n = 2p + 1 \\
z^{1/2} & , & n = 2p \\ 
\ea \right. 
\ee
and $\id_{n-1}$ is the $(n\!-\!1)$-dimensional unit matrix. The centralizer of $(U_1,U_2)$ is $C_{\SU(n)}$, so to obtain an almost commuting rank zero triple on the standard form, with $m = z \cdot \id$, we take any $U_3 \in C_{\SU(n)}$. The order of the triple is here determined by the order of $U_3$ in $\pi_0(\mathcal{Z})$, where $\mathcal{Z}$ is as in section \ref{sec:ModuliSpaceConstruction}. 

To find the spectrum of $(U_1,U_2,U_3)$ we consider a Cartan-Weyl basis of the adjoint representation of the Lie algebra $\su(n)$. Let the Cartan and root generators be given by
\bea 
\label{eqn:CartanSubalgebra}
\{H^i\} & = & \{ \mathcal{E}_{i,i} - \mathcal{E}_{i+1,i+1} \mid  1 \leq i < n \}\\
\label{eqn:RootGenerators}
\{E^{\alpha}\} & = &  \{ \mathcal{E}_{i,j} , \mathcal{E}_{j,i} \mid  1 \leq i < j \leq n \} \,,
\eea
where $\mathcal{E}_{i,j}$ is the $n \times n$ matrix whose only non-zero element is $(\mathcal{E}_{i,j})_{i,j} = 1$. The action, $\ad{U_1}$, of $U_1$ in the adjoint representation, can be diagonalized by first considering the root generators $E^1_{(i)} = \mathcal{E}_{i,1}$ with $i > 1$, whose eigenvalues under $\ad{U_2}$ are $\frac{1}{a_n}(U_2)_{i,i}$, i.e. the $i$:th diagonal element of $U_2$. By repeatedly acting with $U_1$ on $E^1_{(i)}$ we obtain $n-1$ additional root generators, $E^{j+1}_{(i)} =  U_1^j E^1_{(i)} U_1^{-j}$, since $U_1^n = \id$. All $E^{j}_{(i)}$ have the same eigenvalue $\frac{1}{a_n}(U_2)_{i,i}$ under $\ad{U_2}$, since the action of $U_1$ and $U_2$ commute in the adjoint representation. The linear combinations 
\be 
\label{eqn:RootEigenvectorsSU(n)}
T_{(i)} = {\displaystyle \sum_{j=1}^{n}} E^j_{(i)}
\ee
are thus simultaneous eigenvectors of all $\ad{U_i}$ with eigenvalue vectors $(1,\frac{1}{a_n}(U_2)_{i,i},1)$. If the triple is to completely break the gauge group there can be no $(1,1,1)$ eigenvalue vectors. Thus, we find the necessary condition that $m$ in (\ref{eqn:AlmostCommutingPair}) must be a generator of $C_{\SU(n)}$ for $(U_1,U_2,U_3)$ to constitute a rank zero triple. In fact, for $m$ that generate $C_{\SU(n)}$ the pair $(U_1,U_2)$ is unique \cite{Kac:1999,Borel:1999}. The set of generators, $\{c\}$, consists of all elements $c = e^{2 \pi i q / n} \cdot \id$ where $q$ and $n$ are relatively prime. The number of such $q$ is precisely $\varphi(n)$. Thus, from now on we will restrict our attention to triples with $m=c$. By introducing relative phases in the sum (\ref{eqn:RootEigenvectorsSU(n)}) we also obtain all eigenvalues $z^{r=1,\dots,n-1}$ under $\ad{U_1}$. 

Finally, we consider the diagonalization of $\ad{U_1}$ on the Cartan subalgebra generators (\ref{eqn:CartanSubalgebra}), and find the characteristic equation $\mathrm{det}(\ad{U_1} - \la_1) = \sum_{i=0}^{n-1} \la_1^i = 0$, with solutions $\la_1 = z^{r=1,\dots,n-1}$. Since $\ad{U_2}$ is trivial on $H^i$ we thus obtain the eigenvalue vectors $(z^{r=1,\dots,n-1},1,1)$.

All the eigenvalue vectors are non-degenerate and the spectrum of $\SU(n)$ is summarized in Table \ref{tab:SpectrumAlmostCommutingSU(n)} in the form that will be used throughout the rest of this paper. Here, we introduce the notation $\{(\la_1,\la_2,\la_3)\}^{\dag}$ for a set of eigenvalue vectors where $(1,1,1)$ has been excluded. This is convenient since we are only concerned with triples that completely break the gauge group and hence have no such eigenvalues. We note that the $\SL(2,\ZZ)$-invariance of the first two components in the spectrum is manifest.
\begin{table}[!ht]
\begin{center}
\caption{Spectrum for the $m=c$ triples in $\SU(n)$}
\vspace*{2mm}
\begin{tabular}{ccc}
$(\la_1,\la_2,\la_3)$ & degeneracy & $\#$ \\ \hline
\\[-4mm]
\begin{tabular}{ccc}  
$\{(z_1,z_2,1) \mid z_i^n = 1\}^{\dag}$
\end{tabular}
& 
\begin{tabular}{c} 
1
\end{tabular}
&
\begin{tabular}{c}
$n^2-1$
\end{tabular}
\end{tabular}
\label{tab:SpectrumAlmostCommutingSU(n)}
\end{center}
\end{table}

For the $\varphi(n)$ possible $m$ that generate $C_{\SU(n)}$ the moduli space is, according to the above consideration,
\be 
\cM_{\SU(n)}(m) = \overset{n}{\underset{i=1}{\bigcup}} \, \cM_0^{(i)} \,,
\ee 
where the $n$ components correspond to the $n$ inequivalent choices of $U_3$. This result can also be obtained using the previously described method for moduli space construction if $\si$ is taken to be an automorphism that acts simply transitively on the extended Dynkin diagram in Figure \ref{fig:DynkinSu}.

\subsection{The simplest example: SU(2)}
To illustrate the application of the method for determining the spectra of $\cN=4$ supersymmetric Yang-Mills on $T^3$, as presented in the previous section, and also to provide a concrete example of some features of the generic $\SU(n)$ calculation above, we will now pause to treat the simplest case available, that of $G=\SU(2)$, in full detail.

The center of $\SU(2)$, $C_{\SU(2)} \iso \{\id,-\id\}$, is cyclic and contains only one non-trivial element, $-\id$, which is of course also the generator of $C_{\SU(2)}$. Taking $m=-\id$ thus satisfies the criterion above for the existence of almost commuting rank zero triples in the special unitary groups. In fact, there are two such triples, corresponding to the two possible choices of $U_3 \in C_{\SU(2)}$, implying that the moduli space of flat connections for $G=\SU(2)$ and $m=-\id$ is
\be 
\cM_{\SU(2)}(m=-\id) = \cM_0 \, \cup \, \cM'_0 \,.
\ee

The unique choice, up to conjugation, of the first two holonomies is
\be
\ba{ccc}
U_1 = \left( \ba{cc}0 & 1\\-1 & 0\ea \right) 
&,&
U_2 = \left( \ba{cc}i & 0\\0 & -i\ea \right)
\ea
\ee
according to (\ref{eqn:AlmostCommutingPairRepresentation}) and the corresponding Cartan-Weyl basis consist of the three matrices
\be
\ba{ccccc}
H = \left( \ba{cc}1 & 0\\0 & -1\ea \right)
&,&
E^+ = \left( \ba{cc}0 & 1\\0 & 0\ea \right)
&,&
E^- = \left( \ba{cc}0 & 0\\1 & 0\ea \right)
\ea \,.
\ee
The adjoint action of the holonomies on the generators of $\su(2)$ is given in Table \ref{tab:AdjointActionSU(2)}.
\begin{table}[!ht]
\begin{center}
\caption{The adjoint action of the holonomies on the $\su(2)$ generators.}
\vspace*{2mm}
\begin{tabular}{l|ccc}
& $\ad{U_1}$ & $\ad{U_2}$ & $\ad{U_3}$ \\ \hline
$H$ & $-H$ & $H$ & $H$ \\
$E^+$ & $-E^-$ & $-E^+$ & $E^+$ \\
$E^-$ & $-E^+$ & $-E^-$ & $E^-$ \\
\end{tabular}
\label{tab:AdjointActionSU(2)}
\end{center}
\end{table}

Diagonalizing the $\ad{U_i}$-action of the triple $(U_1,U_2,U_3)$ then yields the three eigenvectors and corresponding eigenvalue vectors in Table \ref{tab:SpectrumAlmostCommutingSU(2)}.
\begin{table}[!ht]
\begin{center}
\caption{Spectrum for the $m=-\id$ triples in $\SU(2)$}
\vspace*{2mm}
\begin{tabular}{cc}
$(\la_1,\la_2,\la_3)$ & $T_{\la}$ \\ \hline  
$(-1,1,1)$ & $H$ \\
$(-1,-1,1)$ & $E^+ + E^-$ \\
$(1,-1,1)$ & $E^+ - E^-$ \\
\end{tabular}
\label{tab:SpectrumAlmostCommutingSU(2)}
\end{center}
\end{table}

From these eigenvalue vectors $\vec{\la}$ all eigenvalues of the $i\cD_i$ operator are determined, through the relation $(\om_m)_i = (\mathrm{Arg}\la_i + 2 \pi k_i)$, with $k_i \in \ZZ$, to be
\be
\frac{1}{2\pi} \left( \ba{c} (\om_m)_1 \\ (\om_m)_2 \\ (\om_m)_3 \ea \right) \in \left\{ 
\left( \ba{c} \frac{1}{2} + \ZZ \\ \ZZ \\ \ZZ \ea \right) ,\,
\left( \ba{c} \frac{1}{2} + \ZZ \\ \frac{1}{2} + \ZZ \\ \ZZ \ea \right) ,\,
\left( \ba{c} \ZZ \\ \frac{1}{2} + \ZZ \\ \ZZ \ea \right)
\right\} \,.
\ee
The corresponding angular frequencies $\om_m$, given by (\ref{eqn:AngularFrequency}), completely describes the spectrum at the two isolated vacua of the $\cN=4$ Yang-Mills theory on $T^3$ with gauge group $G=\SU(2)$ through (\ref{eqn:ExpandedTheoryHamiltonian}).

%% file: Spinn.tex
\section{The $\Spin(n)$ Groups}
\label{sec:Spin}
We will begin this section by stating some facts about the spin groups. We will then proceed to calculate the spectra for all rank zero triples, i.e. the triples that break the gauge group completely, in cases where $G = \Spin(n)$.

The spin groups are the simply connected groups obtained through exponentiation of the $\so(n)$ Lie algebra generated by $\Ga^{ij}$. The $\Ga^{ij}$'s are defined as the antisymmetrized product $\Ga^{\mbox{\scriptsize [}i}\Ga^{j\mbox{\scriptsize ]}}$ of matrices $\Ga^i$, obeying the Clifford algebra
\be 
\{\Ga^i,\Ga^j\} = 2\delta^{ij}\id \,.
\ee
There are $\frac{1}{2}(n^2-n)$ independent generators $\Ga^{ij}$ and, unless otherwise indicated, we will take the basis of $\so(n)$ to be
\be
\label{eqn:BasisSo}
\mathcal{B} = \{\Ga^{ij} \mid 1 \leq i < j \leq n\} \,.
\ee
Before proceeding we also note the useful relation 
\be
\exp (\al \Ga^{ij}) = \id \cos(\al) +  \Ga^{ij} \sin(\al) \,
\ee
for the exponentiation of a single generator, obtained using the defining properties of $\Ga^i$.

The center of $G = \Spin(n)$ is
\be 
C = \left\{ 
\ba{lcl}
\{ \id , -\id \} & , &  n = 2q + 1 \\ 
\{ \id , -\id , \Gamma , -\Gamma \} & , &  n = 2q\\
\ea 
\right.
\ee
where $\Ga = \Ga^1 \ldots \Ga^{2q}$ satisfies $(\Ga)^2 = (-1)^q \cdot \id$. This property implies that when $q$ is odd, $\Ga$ generates the center, while the center is not generated by any element for $q$ even. Thus, for $n=4p$ the center is not cyclic, a fact that is reflected in the following isomorphisms for the spin group centers
\be 
C \cong \left\{ 
\ba{lcl}
\ZZ_2 & , &  n = 2p + 1 \\ 
\ZZ_2 \times \ZZ_2 & , &  n = 4p\\
\ZZ_4 & , &  n = 4p + 2
\ea 
\right.
\ee
which are a direct consequence of the order of $\Ga$. The absence of cyclic structure of the center of $\Spin(4p)$, which is a common feature for all other gauge groups, will as previously remarked prevent us from casting some $\hat{m}$-triplets in the standard form using $\SL(3,\ZZ)$-transformations.

The rank zero triples for the various $\Spin(n)$ cases have previously been explicitly constructed in terms of the matrices $\Ga^i$ elsewhere. Therefore, the task of determining their spectra is reduced to simply diagonalizing the action, $\ad{U_i}$, of the holonomies on the basis of the adjoint representation (\ref{eqn:BasisSo}). This is a straightforward but rather tedious process, the result of which is presented below for the rank zero triples in the $\Spin(n)$ groups.

\subsection{The $G = \Spin(2p+1)$ case}
\begin{figure}[!ht]
\begin{center}
\includegraphics[scale=0.3]{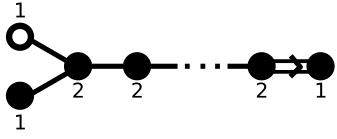}
\caption{Extended Dynkin diagram of $\so(2p+1), p \geq 3$}
\label{fig:DynkinSpin2k1}
\end{center}
\end{figure}

We will begin our consideration of the $\Spin(n)$ groups by studying the case of odd $n$, in which case the extended Dynkin diagram is that shown in Figure \ref{fig:DynkinSpin2k1}.

As can be seen from the diagram for $\Spin(2p+1)$, with $p \geq 3$, the component structure of the moduli space for the two possible $m$-values is
\be 
\cM(m) = \left\{
\ba{lcl}
\cM_p \cup \cM_{p-3} & , & m = \id\\
\cM_{p-1} \cup \cM_{p-2} & , & m = -\id
\ea 
\right. \,.
\ee
From this structure it can be concluded that there is only one $m=\id$ rank zero triple of order $k=2$ in $\Spin(7)$\footnote{The order $k$ of the components are not explicitly indicated in the moduli space component structure. It is however obtained by the method previously described and we will state it for completeness and because it will make a difference in subsequent cases.} and none for $m=-\id$.

\subsubsection{$G = \Spin(7)$}
The treatment of the commuting $\Spin(7)$ triple 
\be
\label{eqn:UCommutingSpin7}
\left\{ \ba{ccl}
U_1 & = & \Ga^1\Ga^2\Ga^3\Ga^4\\
U_2 & = & \Ga^1\Ga^2\Ga^5\Ga^6\\
U_3 & = & \Ga^1\Ga^3\Ga^5\Ga^7
\ea \right. \,,
\ee
given in e.g. \cite{Kac:1999,Henningson:2007a}, is straightforward, the diagonalization yielding the spectrum in Table \ref{tab:SpectrumCommutingSpin7}. 
\begin{table}[!ht]
\begin{center}
\caption{Spectrum for the $k=2$, $m = \id$ triple in $\Spin(7)$}
\vspace*{2mm}
\begin{tabular}{ccc}
$(\la_1,\la_2,\la_3)$ & degeneracy & $\#$ \\ \hline
\\[-4mm]
\begin{tabular}{ccc}
$\{(z_1,z_2,z_3) \mid z_i^2 = 1\}^{\dag}$\\
\end{tabular}
& 
\begin{tabular}{c} 
3
\end{tabular}
&
\begin{tabular}{c}
21
\end{tabular}
\end{tabular}
\label{tab:SpectrumCommutingSpin7}
\end{center}
\end{table}

\subsection{The $G = \Spin(4p)$ case}
\begin{figure}[!ht]
\begin{center}
\includegraphics[scale=0.3]{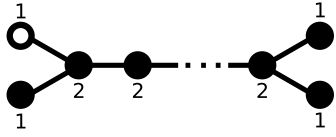}
\caption{Extended Dynkin diagram of $\so(2q), q \geq 4$}
\label{fig:DynkinSpin4k}
\end{center}
\end{figure}

The even spin groups have the extended Dynkin diagrams showed in Figure \ref{fig:DynkinSpin4k}.
Depending on if the number of nodes is odd or even, there are two possible classes of diagram automorphism groups, $\Sigma$, and the construction of the quotient diagram can be divided into two cases. We start with the case where the number is odd and $G = \Spin(4p)$ with $p \geq 2$. By the methods previously described we find the moduli space component structures    
\be 
\label{eqn:ModuliSpaceSpin8}
\cM_{\Spin(8)}(m) = \left\{
\ba{lcl}
\cM_4 \cup \cM_0 & , & m = \id\\
\cM_2 \cup \cM_2' & , & m = -\id,\pm\Ga\\
\cM_1 \cup \cM_1' \cup \cM_0 \cup \cM_0' & , & \hat{m} = (-\id,\Ga,-\Ga)
\ea 
\right.
\ee
and 
\be 
\label{eqn:ModuliSpaceSpin4k}
\cM_{\Spin(4p)}(m) = \left\{
\ba{lcl}
\cM_{2p} \cup \cM_{2p-4} & , & m = \id\\
\cM_{2p-2} \cup \cM_{2p-2}' & , & m = -\id\\
\cM_{p} \cup \cM_{p}' \cup \cM_{p-3} \cup \cM_{p-3}' & , & m = \pm\Ga\\
\cM_{p-1} \cup \cM_{p-1}' \cup \cM_{p-2} \cup \cM_{p-2}' & , & \hat{m} = (-\id,\Ga,-\Ga)
\ea 
\right. \,,
\ee
for $p=2$ and $p \geq 3$ respectively.

The component structure implies that there is one $m=\id$ rank zero triple of order $k=2$ and two $\hat{m}=(-\id,\Ga,-\Ga)$ rank zero triples of order $k=4$ in $\Spin(8)$. Furthermore, there are four almost commuting rank zero triples in $\Spin(12)$, two with $m=\Ga$ and two with $m=-\Ga$, all of order $k=4$.

\subsubsection{$G = \Spin(8)$}
The commuting rank zero triple of $\Spin(8)$, once again taken from e.g. \cite{Kac:1999, Henningson:2007a}, is of the same form as (\ref{eqn:UCommutingSpin7}). Diagonalizing the adjoint action of this triple on the extra eight generators of $\Spin(8)$ extends the spectrum from $\Spin(7)$ to the one in Table \ref{tab:SpectrumCommutingSpin8}. The change is simply an increase from three to four in degeneracy.
\begin{table}[!ht]
\begin{center}
\caption{Spectrum for the $k=2$, $m = \id$ triple in $\Spin(8)$}
\vspace*{2mm}
\begin{tabular}{ccc}
$(\la_1,\la_2,\la_3)$ & degeneracy & $\#$ \\ \hline
\\[-4mm]
\begin{tabular}{ccc}
$\{(z_1,z_2,z_3) \mid z_i^2 = 1\}^{\dag}$\\
\end{tabular}
& 
\begin{tabular}{c} 
4
\end{tabular}
&
\begin{tabular}{c}
28
\end{tabular}
\end{tabular}
\label{tab:SpectrumCommutingSpin8}
\end{center}
\end{table}

The almost commuting triples in $\Spin(8)$ both have $\hat{m} =( -\id,\Ga,-\Ga)$, which can not be put on standard form. One of them is taken from \cite{Henningson:2007b};
\be
\label{eqn:UAlmostCommutingSpin8}
\left\{ \ba{ccl}
U_1 & = & \frac{1}{4}(\id + \Ga^1\Ga^2)(\id - \Ga^3 \Ga^4)(\Ga^5 + \Ga^{6})(\Ga^{7} - \Ga^{8})\\[1mm]
U_2 & = & \frac{1}{4}(\Ga^1 - \Ga^2)(\Ga^3 - \Ga^4)(\id - \Ga^5\Ga^{6})(\id - \Ga^{7}\Ga^{8})\\[1mm]
U_3 & = & \frac{1}{4}(\Ga^1 - \Ga^2)(\Ga^3 + \Ga^4)(\id - \Ga^5\Ga^{6})(\id + \Ga^{7}\Ga^{8})\\
\ea \right. \,.
\ee
By Lemma 12.1.1 in \cite{Borel:1999}, the second one is related to (\ref{eqn:UAlmostCommutingSpin8}) by inverting $U_3$. This operation inverts the $\la_3$ eigenvalue under $\ad{U_3}$, which leaves the spectrum in Table \ref{tab:SpectrumAlmostCommutingSpin8}, obtained by diagonalizing the action of the triple (\ref{eqn:UAlmostCommutingSpin8}), invariant.
\begin{table}[!ht]
\begin{center}
\caption{Spectrum for the $k=4$, $\vec{m}=(-\id, 
\Ga, -\Ga)$ triples in $\Spin(8)$}
\vspace*{2mm}
\begin{tabular}{ccc}
$(\la_1,\la_2,\la_3)$ & degeneracy & $\#$ \\ \hline
\\[-4mm]
\begin{tabular}{ccc}
$\{(z_1,z_2,z_3) \mid z_i^2 = 1,\sum z_i = \pm 1\}^{\dag}$ \\
$\{(\pm i,\pm i,\pm i)\}$
\end{tabular}
& 
\begin{tabular}{c}
2\\2
\end{tabular}
&
\begin{tabular}{c}
12\\\underline{16}
\end{tabular}
\\ 
&&28\\
\end{tabular}
\label{tab:SpectrumAlmostCommutingSpin8}
\end{center}
\end{table}

\subsubsection{$G = \Spin(12)$}
From \cite{Henningson:2007b} we know that the triples
\be
\label{eqn:UAlmostCommutingSpin12}
\left\{ \ba{ccl}
U_1 & = & \frac{1}{2}(\id + \Ga^1\Ga^2)(\id + \Ga^3\Ga^4)\Ga^6\Ga^8\Ga^{10}\Ga^{12}\\[1mm]
U_2 & = & \frac{1}{4}\Ga^2\Ga^4(\Ga^5 - \Ga^6)(\Ga^7 \mp \Ga^8)(\id - \Ga^9\Ga^{10})(\id - \Ga^{11}\Ga^{12})\\[1mm]
U_3 & = & \frac{1}{2}\Ga^3\Ga^4\Ga^7\Ga^8\Ga^9\Ga^{10}
\ea \right.
\ee
in $\Spin(12)$ have rank zero and $m = \pm \Ga$ respectively. Here $\mathcal{K} = \{\id, -\id\}$ which implies that given one of the triples (\ref{eqn:UAlmostCommutingSpin12}) there is an additional conjugacy class obtained through the action of $\Ga$ on $U_3$. Obviously, these two conjugacy classes have the same $m$-value and equivalent actions on the Lie algebra generators. Thus, they possess the same spectrum and correspond to the two rank zero components of the moduli spaces in (\ref{eqn:ModuliSpaceSpin4k}) for $m=\pm\Ga$ respectively. When diagonalizing the adjoint action of one of the possible choices (\ref{eqn:UAlmostCommutingSpin12}), we find the spectrum listed in Table \ref{tab:SpectrumAlmostCommutingSpin12}. From the calculation we also find that the spectra is identical for the other choice, hence Table \ref{tab:SpectrumAlmostCommutingSpin12} shows the spectrum for all isolated points in $\cM_{\Spin(12)}$.

\begin{table}[!ht]
\begin{center}
\caption{Spectrum for the $k=4$, $m=\pm\Ga$ triples in $\Spin(12)$}
\vspace*{2mm}
\begin{tabular}{ccc}
$(\la_1,\la_2,\la_3)$ & degeneracy & $\#$ \\ \hline
\\[-4mm]
\begin{tabular}{ccc}
$\{(z_1,z_2,-1) \mid z_i^2 = 1\}^{\dag}$\\
$\{(z_1,z_2,\pm 1) \mid z_i^4 = 1\}^{\dag}$
\end{tabular}
& 
\begin{tabular}{c} 
1\\2
\end{tabular}
&
\begin{tabular}{c}
4\\ \underline{62}
\end{tabular}
\\
&&66\\
\end{tabular}
\label{tab:SpectrumAlmostCommutingSpin12}
\end{center}
\end{table}

\subsection{The $G = \Spin(4p+2)$ case}
For the $G = \Spin(4p+2)$ groups, with $p \geq 2$, the moduli space of flat connections is found by inspection of the Dynkin diagram in Figure \ref{fig:DynkinSpin4k} to have the following components for the $\SL(3,\ZZ)$-inequivalent $m$-values: 
\be 
\cM(m) = \left\{
\ba{lcl}
\cM_{2p+1} \cup \cM_{2p-3} & , & m = \id\\
\cM_{2p-1} \cup \cM_{2p-1}' & , & m = -\id\\
\overset{4}{\underset{i=1}{\bigcup}} \, \cM_{p-1}^{(i)} & , &  m = \Ga \,.
\ea 
\right.
\ee

From this result we conclude that rank zero triples only appear for the case where $p=1$, i.e. $G = \Spin(6)$. However, $\Spin(6) \cong \SU(4)$ which we have already dealt with in the previous section, and hence there are no new rank zero triples to be found in the $\Spin(4p+2)$ case.

%% file: Spn.tex
\section{The $\Sp(n)$ Groups}
\label{sec:Sp}
\begin{figure}[!ht]
\begin{center}
\includegraphics[scale=0.3]{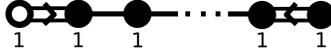}
\caption{Extended Dynkin diagram for $\sp(n)$}
\label{fig:DynkinSp}
\end{center}
\end{figure}

The symplectic groups, $\Sp(n)$, all have centers isomorphic to $\ZZ_2$. This implies that $\Sigma$ contains one non-trivial automorphism of the extended Dynkin diagram, shown in Figure \ref{fig:DynkinSp}.

We find that the moduli space is connected for the case where $n$ is odd and $m$ arbitrary and the case where $n$ is even and $m=\id$. The structure of the moduli space in the remaining case, where $n=2p$ is even and $m=-\id$, is
\be 
\cM_{\Sp(2p)} = \cM_{p} \cup \cM_{p-1} \,.
\ee 
Hence, the only symplectic group admitting a rank zero triple is $\Sp(2)$, which contains precisely one such triple of order $k=2$. To compute the spectrum of this triple we exploit the isomorphism $\Sp(2) \cong \Spin(5)$ to study the problem in the context of $\Spin(5)$ instead. In this group there is an almost commuting rank zero triple \cite{Henningson:2007a}, 
\be
\label{eqn:UAlmostCommutingSpin5}
\left\{ \ba{ccl}
U_1 & = & \Ga^1\Ga^2\\
U_2 & = & \Ga^1\Ga^3\\
U_3 & = & \Ga^5
\ea \right. \,,
\ee
whose spectrum is found to be the one of Table \ref{tab:SpectrumAlmostCommutingSp2}, using the same method as in the previous section.
\begin{table}[!ht]
\begin{center}
\caption{Spectrum for the $k=2$, $m = -\id$ triple in $\Sp(2)$}
\vspace*{2mm}
\begin{tabular}{ccc}
$(\la_1,\la_2,\la_3)$ & degeneracy & $\#$ \\ \hline
\\[-4mm]
\begin{tabular}{ccc}
$\{(z_1,z_2,1) \mid z_i^2 = 1\}^{\dag}$\\
$\{(z_1,z_2,z_3) \mid z_i^2 = 1\}^{\dag}$
\end{tabular}
& 
\begin{tabular}{c} 
1\\1
\end{tabular}
&
\begin{tabular}{c}
3\\\underline{7}
\end{tabular}
\\ & & 10\\
\end{tabular}
\label{tab:SpectrumAlmostCommutingSp2}
\end{center}
\end{table}

%% file: Exceptional.tex
\section{The Exceptional Groups}
\label{sec:exceptional}
It is not possible to deal with the exceptional groups in the same straightforward manner that was used for examining the cases in the two previous sections and hence the treatment in the present section will be somewhat more technical. The general ideas used for computing the spectra for the exceptional groups are to either find $G$ as a subgroup of some other group, where it can be treated explicitly, or to find a subgroup $S \subset G$ and embed almost commuting triples in $S$ in such a way that they break the group $G$ down to $S$, and then break $S$ completely. For all exceptional groups except $G_2$ the latter approach will be convenient. The $\SL(3,\ZZ)$-invariance of the spectra of commuting triples will prove a very useful tool in these considerations.

\subsection{The $G = G_2$ case}
\begin{figure}[!ht]
\begin{center}
\includegraphics[scale=0.3]{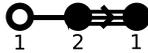}
\caption{Extended Dynkin diagram of $G_2$}
\label{fig:DynkinG2}
\end{center}
\end{figure}

The center, $C_{G_2}$, of $G_2$ is trivial and hence it can not contain any almost commuting triples. The triviality of the center also corresponds to the fact that there are no automorphisms of the extended Dynkin diagram shown in Figure \ref{fig:DynkinG2}.

Applying the general methods, described in \ref{sec:ModuliSpaceConstruction}, for constructing the moduli space to this diagram, we find that the structure of the moduli space is
\be 
\label{eqn:ModuliSpaceG2}
\cM_{G_2} = \cM_2 \cup \cM_0 \,,
\ee 
implying one $m=\id$ rank zero triple of order $\ka=2$ in $G_2$.

In order to find this triple and compute its spectrum we consider $G_2$ as isomorphic to a certain subgroup of $\Spin(7)$, defined according to $G_2 \cong \{g \in \Spin(7) \, | \, g\psi = \psi\} \subset \Spin(7)$, i.e. the stabilizer of $\psi$ in $\Spin(7)$, where $\psi \neq 0$ is a fixed $\Spin(7)$ spinor. 

The commuting rank zero triple (\ref{eqn:UCommutingSpin7}) in $\Spin(7)$ that was considered in the previous section was on the form 
\be
\label{eq:U_G2}
\left\{ \ba{ccl}
U_1 & = & \Ga^1\Ga^2\Ga^3\Ga^4\\
U_2 & = & \Ga^1\Ga^2\Ga^5\Ga^6\\
U_3 & = & \Ga^1\Ga^3\Ga^5\Ga^7
\ea \right. \,,
\ee 
where the Dirac matrices $\Ga^i$ represent the seven dimensional Clifford algebra. Using general methods for explicitly constructing such representations we obtain
\bea
\Ga^{i=1,2,3,4} = \left(\ba{cc}-\hat{\ga}^i & 0\\0 & \hat{\ga}^i\ea\right) & , &
\Ga^5 = \left(\ba{cc}0 & \id_4\\\id_4 & 0\ea\right) \,, \\
\Ga^6 = i \left(\ba{cc}0 & -\id_4\\\id_4 & 0\ea\right) & , &
\Ga^7 = \Ga^1\Ga^2\Ga^3\Ga^4\Ga^5\Ga^6 \,,
\eea
with $\hat{\ga}^1 = \ga^0$ and $\hat{\ga}^{j+1} = i\ga^{j=1,2,3}$, where $\ga^{i=0,1,2,3}$ are the standard four-dimensional Dirac matrices. Acting on a generic $\Spin(7)$ spinor with the holonomies in this representation we find that (\ref{eq:U_G2}) leave a certain spinor $\psi_0$ invariant. Thus, $\{U_i\}$ constitutes a commuting rank zero triple also in $G_2$ \cite{Keurentjes:1998}.

The generators of the $G_2$ Lie algebra in this representation are the linearly independent linear combinations $\sum\Ga^{ij}$ of the $\so(7)$ generators $\{\Ga^{ij} \mid 1 \leq i < j \leq 7\}$ that satisfy $(\sum\Ga^{ij})\psi_0 = 0$, i.e. the basis vectors of the null space of the action of $\Ga^{ij}$ on $\psi_0$. There are $\mbox{dim}(G_2) = 14$ such combinations whose eigenvalues under $\ad{U_i}$ are straightforward to determine using the anticommutation relations of the Dirac matrices, yielding the spectrum in Table \ref{tab:SpectrumCommutingG2}.
\begin{table}[!ht]
\begin{center}
\caption{Spectrum for the $k=2$, $m=\id$ triple in $G_2$}
\vspace*{2mm}
\begin{tabular}{ccc}
$(\la_1,\la_2,\la_3)$ & degeneracy & $\#$ \\ \hline
\\[-4mm]
\begin{tabular}{ccc}
$\{(z_1,z_2,z_3) \mid z_i^2 = 1\}^{\dag}$
\end{tabular}
& 
\begin{tabular}{c} 
2
\end{tabular}
&
\begin{tabular}{c}
14
\end{tabular}
\end{tabular}
\label{tab:SpectrumCommutingG2}
\end{center}
\end{table}

\subsection{The $G = F_4$ case}
\begin{figure}[!ht]
\begin{center}
\includegraphics[scale=0.3]{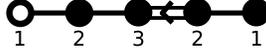}
\caption{Extended Dynkin diagram of $F_4$}
\label{fig:DynkinF4}
\end{center}
\end{figure}

The next to smallest exceptional group, $F_4$, has a trivial center, just like $G_2$. Thus only the extended Dynkin diagram shown in Figure \ref{fig:DynkinF4} has to be considered when constructing $F_4$'s moduli space. 
The resulting component structure is found, using the standard method, to be
\be 
\label{eqn:ModuliSpaceF4}
\cM_{F_4} = \cM_4 \cup \cM_1 \cup \cM_0 \cup \cM '_0 \,,
\ee
where the rank zero components are both of order $k=3$.

Inspecting the diagram and removing one of the $g_{\alpha} = 2$ nodes or the node with $g_{\alpha} = 3$, the possible regular maximal subalgebras are found to be $\mathfrak{s}_1 = \su(4) \oplus \su(2)$ and $\mathfrak{s}_2 = \su(3) \oplus \su(3)$. Consider first $\mathfrak{s}_1$, and the corresponding Lie group
\begin{equation}
S_1 = \SU(2) \times \SU(4) \,.
\end{equation}
The kernel of the embedding map $\iota_1: S_1 \hookrightarrow F_4$ is $K_1 = \{(\id, \id),(-\id, -\id)\}$, and we observe that $\mathfrak{s}_1$ can't be broken completely, since $-\id$ is not a generator of $C_{\SU(4)}$. On the other hand, $\mathfrak{s}_2$, being the tangent space to
\begin{equation}
S_2 = \SU(3) \times \SU(3) \,,
\end{equation}
can be broken completely. Let $c$ (and therefore $c^2$) generate $C_{\SU(3)}$. Under $\iota_2: S_2 \hookrightarrow F_4$, $(\id, \id), (c^2,c)$ and $(c,c^2)$ are mapped to $\id \in F_4$, consequently forming $K_2 = \ker \iota_2 \cong \ZZ_3$. Hence a pair of elements $(U_1,U_2)$ in $\SU(3) \times \SU(3)$ commuting to the non-trivial elements in $K_2$, together with a $U_3$ breaking $F_4$ down to $S_2/K_2$ will constitute a commuting rank zero triple in $F_4$. There are two such choices, $U_3$ and $U'_3 = U_3^2$, corresponding to the two rank zero components in (\ref{eqn:ModuliSpaceF4}). Now considering the decomposition of the adjoint representation of $F_4$, with respect to the subalgebra $\mathfrak{s}_2$, we find
\begin{equation}
\mathbf{52 = (1,8) \oplus (8,1) \oplus (6,\overline{3}) \oplus (\overline{6}, 3)} .
\end{equation}
Here, $\mathbf{3}$ and $\mathbf{8}$ are the vector and adjoint representations of $\SU(3)$ respectively, whereas $\mathbf{6}$ is the symmetric tensor product of two vectors. Taking the result for the first two $\mathbf{(1,8)}$ terms from Table \ref{tab:SpectrumAlmostCommutingSU(n)} and enforcing $\SL(3,\ZZ)$-invariance we find the spectrum listed in Table \ref{tab:SpectrumCommutingF4}. The result is independent of the choice of $U_3$.

\begin{table}[!ht]
\begin{center}
\caption{Spectrum of the $k=3$, $m=\id$ rank zero triples in $F_4$}
\vspace*{2mm}
\begin{tabular}{ccc}
$(\la_1,\la_2,\la_3)$ & degeneracy & $\#$ \\ \hline
\\[-4mm]
\begin{tabular}{ccc}
$\{(z_1,z_2,z_3) \mid z_i^3 = 1\}^{\dag}$
\end{tabular}
& 
\begin{tabular}{c} 
2
\end{tabular}
&
\begin{tabular}{c}
52
\end{tabular}
\end{tabular}
\label{tab:SpectrumCommutingF4}
\end{center}
\end{table}

\subsection{The $G = E_6$ case}
\begin{figure}[!ht]
\begin{center}
\includegraphics[scale=0.3]{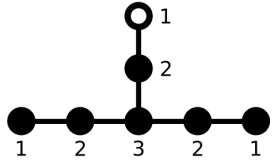}
\caption{Extended Dynkin diagram of $E_6$}
\label{fig:DynkinE6}
\end{center}
\end{figure}

By inspection of the extended Dynkin diagram in Figure \ref{fig:DynkinE6} of $E_6$, whose center is $C_{E_6} \iso \ZZ_3$, the structure of $\mathcal{M}_{E_6}(m)$ is found to be
\be
\label{eqn:ModuliSpaceE6}
\mathcal{M}_{E_6}(m) = \left\{
\ba{lcl}
\mathcal{M}_6 \cup \mathcal{M}_2 \cup \mathcal{M}_0 \cup \mathcal{M}'_0  & , & m = \id \\
\mathcal{M}_2 \cup \mathcal{M}'_2 \cup \mathcal{M}''_2 \cup \mathcal{M}_0 \cup \mathcal{M}'_0 \cup \mathcal{M}''_0 & , & m = e^{\pm 2\pi i /3} \cdot \id  \,.
\ea 
\right.
\ee
The two rank zero $m=\id$ components have order $k=3$ while two of the isolated $m = e^{\pm 2\pi i /3} \cdot \id$ components have $k=6$ and one have $k=2$.

The maximal subalgebras of $E_6$ are $\mathfrak{s}_1 = \mathfrak{su}(3) \oplus \mathfrak{su}(3)  \oplus \mathfrak{su}(3)$ and $\mathfrak{s}_2 = \mathfrak{su}(6) \oplus \mathfrak{su}(2)$, which is also seen from the extended Dynkin diagram by removing the node with dual Coxeter label $g_{\alpha}=3$ or any of the three with $g_{\alpha}=2$. The subgroup $S_1 \subset E_6$ corresponding to three $\mathfrak{su}(3)$ direct summands, 
\begin{equation}
S_1 = \mathrm{SU(3)} \times \mathrm{SU(3)}  \times \mathrm{SU(3)}/K_1 \subset E_6 \,,
\end{equation} 
contains two commuting triples with zero rank of order $k=3$. Here, the kernel is $K_1 = \{(c^k,c^k,c^k) \mid k = 0,1,2\} \cong \ZZ_3$ and $c$ is one of the generators of $C_{SU(3)}$. Thus by choosing the same almost commuting pair in each $\mathrm{SU(3)}$ factor, their direct product commutes in $E_6$ and we get two possible, but equivalent, choices for $U_1, U_2 \in E_6$. 

Consider $\mathfrak{s}_1$, and let $\mathbf{8}$ and $\mathbf{3}$ denote the adjoint and vector representations of $\mathfrak{su}(3)$ respectively. Then the decomposition of $\mathbf{78}$, the adjoint of $E_6$, becomes 
\begin{equation}
\mathbf{78 = (8,1,1) \oplus (1,8,1) \oplus(1,1,8) \oplus (3,3,3) \oplus(\overline{3},\overline{3},\overline{3})} .
\end{equation}
\noindent The center of $E_6$, i.e. $C_{E_6} \cong \mathbb{Z}_3$ is represented by $\{(\id,\id,\id),(\id, c^2, c),(\id,c,c^2)\} \in S_1$. Note that $S_1$ does not contain an almost commuting triple as all $\mathfrak{su}(3)$ factors can't be broken simultaneously if $U_1$ and $U_2$ commute to a non-trivial center element in $E_6$. For the commuting triple to break the generators in $E_6$ not contained in $\mathfrak{s}_1$ we choose $U_3 = (c^k, c^l, c^m)$ as a direct product of center elements in $C_{SU(3)}$ and require that $U_3$ is not in the center of $E_6$ \cite{Witten:2000}. Such elements have $k+l+m \neq 0 \ \mathrm{mod} \ 3$ and as $\mathcal{K} = C_G$ for commuting triples there are only two such inequivalent such choices $U_3$ and $U'_3$ after dividing out $K_1$ and after invoking the action of the center. The two choices can be related as $U'_3 = U_3^2$. As indicated by (\ref{eqn:ModuliSpaceE6}) there are no other isolated triples for $m=\id$. Consider now diagonalizing $\mathfrak{s}_1$ under the adjoint action of $U_1$, $U_2$, and $U_3$. 

The contribution from the $\mathbf{(8,1,1)}$ direct summand and its two permutations to the spectrum is taken from section \ref{sec:Su} to be the eigenvalue vectors $(z_1,z_2,1)$ with $z_i^3 =1$ and three-fold degeneracy. The $\SL(3,\ZZ)$-invariance then gives the eigenvalue vectors with third roots of unity as $\lambda_3$, resulting in the spectrum listed in Table \ref{tab:SpectrumCommutingE6}. Again the spectra is independent of the conjugacy class.

\begin{table}[!ht]
\begin{center}
\caption{Spectrum for the $k=3$, $m=\id$ triples in $E_6$}
\vspace*{2mm}
\begin{tabular}{ccc}
$(\la_1,\la_2,\la_3)$ & degeneracy & $\#$ \\ \hline
\\[-4mm]
\begin{tabular}{ccc}
$\{(z_1,z_2,z_3) \mid z_i^3 = 1\}^{\dag}$ 
\end{tabular}
& 
\begin{tabular}{c} 
3
\end{tabular}
&
\begin{tabular}{c}
78
\end{tabular}
\end{tabular}
\label{tab:SpectrumCommutingE6}
\end{center}
\end{table}

Next, we consider $\mathfrak{s}_2$ and embed its exponentiation as a subgroup of $E_6$,
\begin{equation}
S_2 = \SU(2) \times \SU(6) /K_2 \ \subset E_6
\end{equation}
Here $K_2 \cong \ZZ_2$ consists of the elements $(\id,\id)$ and $(-\id,-\id)$ and we represent the center of $E_6$ with the set $\{ (\id,\id), (-\id, c), (-\id, c^5)\} \in S_2$, where $c$ generates $C_{\SU(6)}$. As we recall from section \ref{sec:Su}, $U_1, U_2 \in \SU(6)$ satisfying equation (\ref{eqn:AlmostCommutingPair}) is not rank zero for $m = -\id$. Hence $S_2$ contains no commuting triple with finite centralizer. On the other hand, $S_2$ contains almost commuting triples. The non-trivial elements in $C_{E_6}$, viewed as elements of $S_2$, consist of two generators and hence we choose the standard irreducible pairs for $\mathrm{SU}(2)$ and $\mathrm{SU}(6)$ to constitute $U_1$ and $U_2$. For the $m = e^{\pm 2\pi i /3} \cdot \id$ triples in $E_6$, $\mathcal{K} =\id$ and for $U_3$ we then have three possible choices, all of which are related through the action of the non-trivial elements of the center and hence correspond to the three isolated components of $\mathcal{M}_{E_6}(m \neq \id)$. The decomposition of $\mathbf{78}$ under $\mathfrak{s}_2$ is
\begin{equation}
\mathbf{78 = (3,1) \oplus (1,35) \oplus (2,20)} .
\end{equation}
where $\mathbf{3}$ and $\mathbf{2}$ are the adjoint and vector representations of $\mathfrak{su}(2)$ and $\mathbf{35}$ is the adjoint of $\mathfrak{su}(6)$. Here, $\mathbf{20}$ is the totally antisymmetric product of three $\mathfrak{su}(6)$ vectors. Diagonalizing the adjoint actions $\ad{U_i}$, the spectral contribution from the first two summands is as in Table \ref{tab:SpectrumAlmostCommutingSU(n)}. Considering the last summand $\mathbf{(2,20)}$, the three possible $U_3$ all act as $-\id$. Finding the eigenvalues of $\ad{U_1}$ and $\ad{U_2} $ in this case is a straightforward calculation and the complete spectrum is listed in Table \ref{tab:SpectrumAlmostCommutingE6}. As the choices of $U_3$ differ only up to the left action of a center element, the spectrum is identical for the three conjugacy classes.

\begin{table}[!ht]
\begin{center}
\caption{Spectrum for the $k=2$ and $k=6$, $m=e^{\pm 2\pi i /3} \cdot \id$ triples in $E_6$}
\vspace*{2mm}
\begin{tabular}{ccc}
$(\la_1,\la_2,\la_3)$ & degeneracy & $\#$ \\ \hline
\\[-4mm]
\begin{tabular}{ccc}
$\{(z_1,z_2,z_3) \mid z_i^2=1\}^{\dag}$ \\
$\{(z_1,z_2,\pm 1) \mid z_i^6 = 1\}^{\dag}$
\end{tabular}
& 
\begin{tabular}{c} 
1 \\
1
\end{tabular}
&
\begin{tabular}{r}
7 \\ \underline{71}
\end{tabular}
\\& & 78\\
\end{tabular}
\label{tab:SpectrumAlmostCommutingE6}
\end{center}
\end{table}

\subsection{The $G = E_7$ case}
The component structures (\ref{eqn:ModuliSpaceE7Identity}) and (\ref{eqn:ModuliSpaceE7NegativeIdentity}) of $\cM_{E_7}$, allow us to infer that there are two $m=\id$ triples and four $m=-\id$ triples in $E_7$ of rank zero. The $m=\id$ triples are of order $k=4$ while among the $m=-\id$ triples two are of order $k=3$ and two are of order $k=6$. To find them we proceed in the same way as in the previous case, by studying the maximal regular subalgebras of $E_7$ found by removing nodes in the extended Dynkin diagram. Removing the node with $g_{\al} = 4$ we obtain $\su(4) \oplus \su(4) \oplus \su(2)$ which can in turn be embedded in $\mathfrak{s}_1 = \su(2) \oplus \so(12)$, obtained by removing either of the two $g_{\al} = 2$ nodes related by $\Sigma$. By removing any one of the nodes with $g_{\al} = 3$, or the top $g_{\al} = 2$ one, the remaining subalgebras $\mathfrak{s}_2 = \su(3) \oplus \su(6)$ and $\mathfrak{s}_3 = \su(8)$ are obtained.

The subgroup corresponding to $\mathfrak{s}_1$ is
\be
\label{eqn:S1E7}
S_1 = \SU(2) \times \Spin(12) / K_1 \subset E_7 \,,
\ee
where $K_1 = \{(\id,\id),(-\id,\Ga)\}$. The center of $E_7$ is thus represented by the set of elements $\{(\id,\id),(\id,-\Ga)\} \in S_1$. In order to completely break each factor of $S_1$ simultaneously a triple $(U_1,U_2,U_3)$ in $E_7$ must have an $\SU(2)$ part with $m=-\id$. Breaking the $\Spin(12)$ factor in such a way as to yield $m=\Ga$ thus produces a triple in $S_1$ with $m=(-\id,\Ga)\sim(\id,\id)$ in $E_7$. In order to break the $E_7$ generators that are not in $\mathfrak{s}_1$ the third element of the triple is taken to be $U_3 = (c,c') \notin C_{E_7}$ where $c\in C_{\SU(2)}$ and $c'\in C_{\Spin(12)}$. There are four possible choices of $U_3$, related pairwise through the action of the center $C_{E_7}$. The two inequivalent ones are related through $U'_3 = U_3^l$, with $l=3$, and hence represent the two conjugacy classes of commuting triples in $E_7$, exhausting the $m=\id$ rank zero components of (\ref{eqn:ModuliSpaceE7Identity}). Before proceeding to compute the spectra of these commuting triples, we note that it is not possible to embed an almost commuting triple in $S_1$ since the only other possibility for breaking the $\Spin(12)$ part has $m=-\Ga$, which does not yield a composite $m$-value in $C_{E_7}$. 

Under the $S_1$ subgroup the adjoint representation of $E_7$ is decomposed according to
\be
\label{eqn:S1DecompE7}
\mathbf{133 = (3,1) \oplus (1,66) \oplus (2,32')}
\ee
where $\mathbf{2}$ and $\mathbf{3}$ are the vector and adjoint representations of $\SU(2)$, and $\mathbf{66}$ and $\mathbf{32'}$ are the adjoint and anti-chiral spinor representations of $\Spin(12)$, respectively. The contributions to the spectrum of eigenvalue vectors from the first two terms in (\ref{eqn:S1DecompE7}) have already been computed. Combined, they are $\{(z_1,z_2,z_3) \mid z_i^2=1\}^{\dag}$, without degeneracy, and $\{(z_1,z_2,\pm 1) \mid z_i^4=1\}^{\dag}$, with 2-fold degeneracy. The $\SL(3,\ZZ)$-invariance of the spectra for commuting triples then implies the full spectrum in Table \ref{tab:SpectrumCommutingE7}. We note that this is independent of the choice of $U_3$.
\begin{table}[!ht]
\begin{center}
\caption{Spectrum for the $k=4$, $m=\id$ triples in $E_7$}
\vspace*{2mm}
\begin{tabular}{ccc}
$(\la_1,\la_2,\la_3)$ & degeneracy & $\#$ \\ \hline
\\[-4mm]
\begin{tabular}{ccc}
$\{(z_1,z_2,z_3) \mid z_i^2 = 1\}^{\dag}$\\
$\{(z_1,z_2,z_3) \mid z_i^4 = 1\}^{\dag}$ 
\end{tabular}
& 
\begin{tabular}{c} 
1\\2
\end{tabular}
&
\begin{tabular}{r}
7\\ \underline{126}\\
\end{tabular}
\\ & & 133\\
\end{tabular}
\label{tab:SpectrumCommutingE7}
\end{center}
\end{table}

The next subalgebra to consider, $\mathfrak{s}_2$, has a corresponding subgroup
\be
\label{eqn:S2E7}
S_2 = \SU(3) \times \SU(6) / K_2 \subset E_7 \,,
\ee
with $K_2 = \{(\id,\id),(c^2,c^4),(c^4,c^2)\}$, where $c$ is taken to be a generator of $C_{\SU(6)}$. Here, we exploit the fact that the center of $\SU(3)$ can be expressed in terms of $c$ as $\{\id,c^2,c^4\}$. The center of $E_7$ is represented by the elements $\{(\id,\id),(\id,c^3)\} \in S_2$. To simultaneously break both $\SU$-factors completely, $U_1,U_2 \in E_7$ must commute to $(c_{12},c_{12}') \in C_{E_7}$ where $c_{12}$ and $c_{12}'$ generate $C_{\SU(3)}$ and $C_{\SU(6)}$ respectively. There are two possible combinations both of which are identified with $(\id,c^3)$ in $S_2$ and hence there are no commuting rank zero triples of $E_7$ in $S_2$. However, the possibility to construct an almost commuting triple by adding $U_3 = (c_3,c_3')$, where $c_3 \in C_{\SU(3)}$ and $c_3' \in C_{\SU(6)}$, to one choice of $U_1,U_2$ remains. In doing so we must demand that $U_3 \notin C_{E_7}$ in order to break the remaining $E_7$ generators. There are then only four choices of $U_3 \in S_2$, represented by $\{(\id,c),(\id,c^2),(\id,c^4),(\id,c^5)\}$. The stabilizer of the induced action of the center on the conjugacy classes of almost commuting rank zero triples is $\mathcal{K} = \id$, i.e. the action of $C_{E_7} \ni \ga = (\id,c^3)$ on $U_3$ is non-trivial on the set of conjugacy classes. This implies that $(\id,c)$ and $(\id,c^4)$, which are related through this action, represent distinct conjugacy classes. Similarly, $(\id,c^2)$ and $(\id,c^5)\}$ are also related through the action of $\ga$ and thus correspond to distinct conjugacy classes. Finally, all other possible conjugacy relations are ruled out using the fact that all conjugacy classes are of order $k=3$ or $k=6$. Hence, we can infer that the four possible choices of $U_3$ produces almost commuting triples that are inequivalent under conjugation and correspond to the four rank zero components of the moduli space (\ref{eqn:ModuliSpaceE7NegativeIdentity}) for $m=-\id$ in $E_7$. 

The decomposition of the adjoint representation under the subgroup $S_2$ is
\be
\label{eqn:S2DecompE7}
\mathbf{133 = (8,1) \oplus (1,35) \oplus (3,\overline{15}) \oplus (\overline{3},15)}
\ee
where $\mathbf{3}$ is the vector representation of $\SU(3)$ while $\mathbf{8}$ and $\mathbf{35}$ are the adjoint representations of the two $\SU$-factors. The $\mathbf{15}$ is the antisymmetric product of two fundamental $\mathbf{6}$'s of $\SU(6)$. From Table \ref{tab:SpectrumAlmostCommutingSU(n)} the contribution to the spectrum of eigenvalue vectors from the first two summands is found to be $\{(z_1,z_2,z_3) \mid z_i^3=1\}^{\dag}$ and $\{(z_1,z_2,z_3) \mid z_i^6=1\}^{\dag}$ where both sets are non-degenerate. 

Using the construction of $\mathbf{15}$ from the fundamental $\mathbf{6}$, diagonalizing $\ad{U_1}$ and $\ad{U_2}$ on the two last summands is found to produce the same set of eigenvalue vector components $(\la_1,\la_2)$; $\{(z_1,z_2) \mid z_i^3=1\}$ and $\{(z_1,z_2) \mid z_i^6=1\}$ without degeneracy. It thus remains to consider the eigenvalues $\la_3$ under $\ad{U_3}$, whose action on $\mathbf{(3,\overline{15})}$ and $\mathbf{(\overline{3},15)}$ is $c_3\bar{c}_3'^2$ and $\bar{c}_3c_3'^2$. Considering the four inequivalent $U_3$'s, we find the complete spectrum in Table \ref{tab:SpectrumAlmostCommutingE7}, which once again is independent of the choice of $U_3$.  

\begin{table}[!ht]
\begin{center}
\caption{Spectrum for the $k=3$ and $k=6$, $m=-\id$ triples in $E_7$}
\vspace*{2mm}
\begin{tabular}{ccc}
$(\la_1,\la_2,\la_3)$ & degeneracy & $\#$ \\ \hline
\\[-4mm]
\begin{tabular}{ccc}
$\{(z_1,z_2,z_3) \mid z_i^3 = 1\}^{\dag}$\\ 
$\{(z_1,z_2,z_3) \mid z_{1,2}^6 = 1\,, z_3^3 = 1\}^{\dag}$
\end{tabular}
& 
\begin{tabular}{c} 
1\\1
\end{tabular}
&
\begin{tabular}{r}
26\\ \underline{107}\\
\end{tabular}
\\ & & 133 \\
\end{tabular}
\label{tab:SpectrumAlmostCommutingE7}
\end{center}
\end{table}

Finally, considering the subalgebra $\mathfrak{s}_3$, we note that it is not possible to embed a commuting or almost commuting triple in a subgroup $S_3 = \SU(8) / K_3$, with $K_3 = \{\id,c^4\}$ and $c$ a generator of $C_{\SU(8)}$, of $E_7$ in such way that all $\SU(8)$ generators are broken. This is due to the fact that no generators of $C_{\SU(8)}$ are identified with elements of $C_{E_7}$, represented by $\{\id,c^2\} \in S_3$, when dividing out the subgroup $K_3$. There are hence no additional rank zero triples in $S_3$, which agrees with the counting of rank zero components of the moduli spaces (\ref{eqn:ModuliSpaceE7Identity}) and (\ref{eqn:ModuliSpaceE7NegativeIdentity}).  

\subsection{The $G = E_8$ case}
\begin{figure}[!ht]
\begin{center}
\includegraphics[scale=0.3]{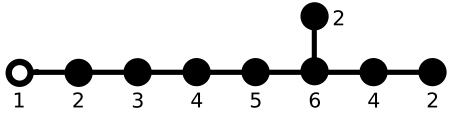}
\caption{Extended Dynkin diagram of $E_8$}
\label{fig:DynkinE8}
\end{center}
\end{figure}

The last remaining, and also the largest, exceptional group is $E_8$, whose center is trivial, implying that it contains no almost commuting triples. This is also reflected in the lack of diagram automorphisms in the extended Dynkin diagram in Figure \ref{fig:DynkinE8}.

The moduli space of commuting, $m=\id$, triples can be determined from the diagram in Figure \ref{fig:DynkinE8} to have the components
\be
\label{eqn:ModuliSpaceE8}
\mathcal{M}_{E_8} = \mathcal{M}_8 \cup \mathcal{M}_4 \cup \mathcal{M}_2 \cup \mathcal{M}'_2 \cup \mathcal{M}_1 \cup \mathcal{M}'_1 \,\overset{6}{\underset{i=1}{\bigcup}} \, \mathcal{M}^{(i)}_0\,,
\ee
from which we conclude that there are six commuting rank zero triples in $E_8$. Out of these six triples four have order $k=5$ and two have order $k=6$. As we will shortly see this is the only case where the spectra of triples with the same $m$-value depend on the order $k$. As for previous cases we proceed by studying the maximal regular subalgebras of $E_8$. By removing nodes in the extended Dynkin diagram we find $\mathfrak{s}_1=\su(5)\oplus\su(5)$, $\mathfrak{s}_2=\su(3)\oplus E_6$, $\mathfrak{s}_3=\su(2)\oplus E_7$, $\mathfrak{s}_4=\su(9)$ and $\mathfrak{s}_5=\so(16)$. When removing some of the nodes we obtain additional subalgebras, but they can all be embedded in some of the previous five, which therefore constitute the maximal subalgebras.

It is straightforward to show that it is not possible to embed any commuting triples in either of the subgroups corresponding to $\mathfrak{s}_3$, $\mathfrak{s}_4$ or $\mathfrak{s}_5$ in such a way that all $E_8$ generators are broken. Thus we proceed to consider the two remaining subalgebras $\mathfrak{s}_1$ and $\mathfrak{s}_2$.

The subgroup corresponding to $\mathfrak{s}_1$ is
\be
\label{eqn:S1E8}
S_1 = \SU(5) \times \SU(5) / K_1 \subset E_8
\ee
where $K_1 = \{(\id,\id),(c,c^3),(c^2,c),(c^3,c^4),(c^4,c^2)\}$, with $c$ a generator of $C_{\SU(5)}$. In order to break all generators of the $\SU$-factors, the first two holonomies $U_1,U_2\in E_8$ must commute to $(c_{12},c'_{12})$ where $c_{12}$ and $c'_{12}$ must generate $C_{\SU(5)}$. We must of course also demand that $(c_{12},c'_{12}) \sim (\id,\id)$ for the triple to be commuting. To break the remaining generators we must then take $U_3 = (c_3,c'_3) \notin K_1$, where $c_3,c'_3 \in C_{\SU(5)}$. There are four possible such elements, represented by the elements $\{(\id,c),(\id,c^2),(\id,c^3),(\id,c^4)\} \in S_1$, which can all be expressed as $(\id,c)^l$ with $l=1,2,3,4$, i.e. with $l$ coprime to $k=5$. From section \ref{sec:ConjugacyClasses} it is then immediately clear that the choices of $U_3$ correspond to the four distinct $k=5$ rank zero components of the moduli space (\ref{eqn:ModuliSpaceE8}).

Under $S_1$, the adjoint representation of $E_8$, which coincides with the fundamental representation, decomposes according to
\be 
\label{eqn:H1DecompE8}
\mathbf{248 = (24,1) \oplus (1,24) \oplus (10,5) \oplus (\overline{10},\overline{5}) \oplus (5,\overline{10}) \oplus (\overline{5},10)} \,,
\ee
where $\mathbf{5}$ and $\mathbf{24}$ are the fundamental and adjoint $\SU(5)$ representations and $\mathbf{10}$ is the antisymmetric product of two $\mathbf{5}$'s. The spectral contribution from the first two terms we obtain from Table \ref{tab:SpectrumAlmostCommutingSU(n)} as the eigenvalue vectors $\{(z_1,z_2,1) \mid z_i^5=1\}^{\dag}$ with 2-fold degeneracy. Using the $\SL(3,\ZZ)$-invariance of the spectrum yields the full spectrum in Table \ref{tab:SpectrumCommutingE8K5}, which is once again independent on the specific choice of the third $k=5$ holonomy $U_3$. 
\begin{table}[!ht]
\begin{center}
\caption{Spectrum for the $k=5$, $m=\id$ rank zero triples in $E_8$}
\vspace*{2mm}
\begin{tabular}{ccc}
$(\la_1,\la_2,\la_3)$ & degeneracy & $\#$ \\ \hline
\\[-4mm]
\begin{tabular}{ccc}
$\{(z_1,z_2,z_3) \mid z_i^5 = 1\}^{\dag}$
\end{tabular}
& 
\begin{tabular}{c} 
2
\end{tabular}
&
\begin{tabular}{r}
248
\end{tabular}
\end{tabular}
\label{tab:SpectrumCommutingE8K5}
\end{center}
\end{table}

The subgroup corresponding to the last remaining maximal regular subalgebra $\mathfrak{s}_2$ is
\be
\label{eqn:S2E8}
S_2 = \SU(3) \times E_6 / K_2 \subset E_8 \,.
\ee
The centers of the two factor groups are isomorphic, $C_{\SU(3)} \cong C_{E_6} \cong \ZZ_3$. The subgroup divided out in order to accommodate $S_2$ in $E_8$ is $K_2=\{(\id,\id),(c,c^2),(c^2,c)\}$ where $c$ is a generator of $\ZZ_3$. In order to completely break the $\SU(3)$ generators, $U_1,U_2 \in E_8$ must commute to one of the non-trivial elements of $K_2$. These elements both allow the $E_6$ generators to be completely broken by taking rank zero $E_6$ components. To break the remaining generators we require that $U_3 = (c_3,c'_3)$, where $c_3 \in C_{\SU(3)}$ and $c'_3 \in C_{E_6}$, is not identified by the identity element of $E_8$. There are two choices of $U_3$, represented by the elements $\{(\id,c),(\id,c^2)\} \in S_2$, not identified by division of $K_2$. They are related through $(\id,c^2) = (\id,c)^l$, with $l = 5$ and are, according to the discussions in section \ref{sec:ConjugacyClasses}, thus representatives of the two $k=6$ rank zero components of the moduli space. That both triples have order $k=6$ can be seen by considering the subalgebra $\su(2) \oplus \su(3) \oplus \su(6)$, obtained by removing the node with $g_{\al}=6$ in the extended Dynkin diagram of $E_8$, which can be embedded in $\mathfrak{s}_2$.

Under $S_2$, the adjoint $\mathbf{248}$ representation of $E_8$ is decomposed according to
\be
\label{eqn:H2DecompE8}
\mathbf{248 = (8,1) \oplus (1,78) \oplus (3,27) \oplus (\overline{3},\overline{27})}
\ee
where $\mathbf{3}$, $\mathbf{27}$ and $\mathbf{8}$, $\mathbf{78}$ are the fundamental and adjoint representations of $\SU(3)$ and $E_6$ respectively. The contributions from the two first summands in (\ref{eqn:H2DecompE8}) have been computed previously, Tables \ref{tab:SpectrumAlmostCommutingSU(n)} and \ref{tab:SpectrumAlmostCommutingE6}, to be the eigenvalue vectors $\{(z_1,z_2,z_3) \mid z_i^2=1\}^{\dag}$, $\{(z_1,z_2,1) \mid z_i^3=1\}^{\dag}$ and $\{(z_1,z_2,\pm 1) \mid z_i^6=1\}^{\dag}$ without degeneracy. Using again the $\SL(3,\ZZ)$-invariance of spectra of commuting triples we obtain the spectrum in Table \ref{tab:SpectrumCommutingE8K6} for the two $k=6$ triples.
\begin{table}[!ht]
\begin{center}
\caption{Spectrum for the $k=6$, $m=\id$ rank zero triples in $E_8$}
\vspace*{2mm}
\begin{tabular}{ccc}
$(\la_1,\la_2,\la_3)$ & degeneracy & $\#$ \\ \hline
\\[-4mm]
\begin{tabular}{ccc}
$\{(z_1,z_2,z_3) \mid z_i^2 = 1\}^{\dag}$\\
$\{(z_1,z_2,z_3) \mid z_i^3 = 1\}^{\dag}$\\
$\{(z_1,z_2,z_3) \mid z_i^6 = 1\}^{\dag}$\\ 
\end{tabular}
& 
\begin{tabular}{c} 
1\\1\\1
\end{tabular}
&
\begin{tabular}{r}
7\\26\\\underline{215}\\
\end{tabular}
\\ & &248\\
\end{tabular}
\label{tab:SpectrumCommutingE8K6}
\end{center}
\end{table}

%% file: Summary.tex
\section{Results and discussion}
\renewcommand\arraystretch{1.5}
In this section we will compile the results obtained in sections \ref{sec:Su}-\ref{sec:exceptional} and discuss their interpretation. The commuting and almost commuting triples are presented separately, in tables \ref{tab:ResultSummaryCommuting} and \ref{tab:ResultSummaryAlmostCommuting} respectively, due to certain qualitative difference e.g. transformation properties under the mapping class group $\SL(3,\ZZ)$. Furthermore, we tabulate the set of eigenvalue vectors $\{(\la_1,\la_2,\la_3)\}$ rather than the spectrum of the Hamiltonian to retain manifest all the structure of the results. Recall from section \ref{sec:LowEnergyEffectiveTheories} that the connection to the physical spectrum is made through the relation
\be
(\om_m)_i = (\mathrm{Arg} \la_i + 2\pi k_i) \,,
\ee
with $k_i \in \ZZ$, between the eigenvalue vectors and the eigenvalues $(\om_m)_i$ of the differential operator $i\cD_i$.

We will adhere to the notation used in the previous chapters, denoting by $\{(\la_1,\la_2,\la_3)\}^{\dag}$ a set of eigenvalue vectors where $(1,1,1)$ is excluded. The order of the $\SU(n)$ triples are determined by the order $k$ of the third holonomy, $U_3$, in $\mathcal{Z}$ as discussed in section \ref{sec:Su}.

\begin{table}[!ht]
\begin{center}
\caption{Spectra for commuting rank zero triples}
\begin{tabular}{|c|c|c|}
\hline
$G$ & $k$ & $(\la_1,\la_2,\la_3)$ [Deg.] \\ \hline\hline

$G_2$ & 2 &
\begin{tabular}{c}
$\{(z_1,z_2,z_3) \mid z_i^2 = 1\}^{\dag}$\\
\end{tabular}
\begin{tabular}{c}
$[2]$
\end{tabular} \\ \hline

$Spin(7)$ & 2 &
\begin{tabular}{c}
$\{(z_1,z_2,z_3) \mid z_i^2 = 1\}^{\dag}$\\
\end{tabular}
\begin{tabular}{c}
$[3]$
\end{tabular} \\ \hline

$Spin(8)$ & 2 &
\begin{tabular}{c}
$\{(z_1,z_2,z_3) \mid z_i^2 = 1\}^{\dag}$\\
\end{tabular}
\begin{tabular}{c}
$[4]$
\end{tabular} \\ \hline

$F_4$ & 3 &
\begin{tabular}{c}
$\{(z_1,z_2,z_3) \mid z_i^3 = 1\}^{\dag}$\\
\end{tabular}
\begin{tabular}{c}
$[2]$
\end{tabular} \\ \hline

$E_6$ & 3 &
\begin{tabular}{c}
$\{(z_1,z_2,z_3) \mid z_i^3 = 1\}^{\dag}$\\
\end{tabular}
\begin{tabular}{c}
$[3]$
\end{tabular} \\ \hline

$E_7$ & 4 &
\begin{tabular}{c}
$\{(z_1,z_2,z_3) \mid z_i^2 = 1\}^{\dag}$\\
$\{(z_1,z_2,z_3) \mid z_i^4 = 1\}^{\dag}$
\end{tabular}
\begin{tabular}{c}
$[1]$ \\ $[2]$
\end{tabular} \\ \hline

$E_8$ & 5 &
\begin{tabular}{c}
$\{(z_1,z_2,z_3) \mid z_i^5 = 1\}^{\dag}$\\
\end{tabular}
\begin{tabular}{c}
$[2]$
\end{tabular} \\ \hline

$E_8$ & 6 &
\begin{tabular}{c}
$\{(z_1,z_2,z_3) \mid z_i^2 = 1\}^{\dag}$\\
$\{(z_1,z_2,z_3) \mid z_i^3 = 1\}^{\dag}$\\
$\{(z_1,z_2,z_3) \mid z_i^6 = 1\}^{\dag}$
\end{tabular}
\begin{tabular}{c}
$[1]$ \\ $[1]$ \\ $[1]$
\end{tabular} \\ \hline

\end{tabular}
\label{tab:ResultSummaryCommuting}
\end{center}
\end{table}

\begin{table}[!ht]
\begin{center}
\caption{Spectra for almost commuting triples}
\vspace*{2mm}
\begin{tabular}{|c|c|c|c|}
\hline
$G$ & $m (\vec{m})$  & $k$ & $(\la_1,\la_2,\la_3)$ [Deg.] \\ \hline \hline

$\SU(n)$ & $\{c\}$ & - &
\begin{tabular}{c}
$\{(z_1,z_2,1) \mid z_i^n = 1\}^{\dag}$
\end{tabular}
\begin{tabular}{c}
$[1]$
\end{tabular}
\\ \hline

$\Spin(8)$ & $(-\id,\Ga,-\Ga)$ & 4 &
\begin{tabular}{c}
$\{(z_1,z_2,z_3) \mid z_i^2 = 1 , \sum z_i = \pm 1\}^{\dag}$\\
$\{(\pm i,\pm i,\pm i)\}^{\dag}$
\end{tabular}
\begin{tabular}{c}
$[2]$ \\ $[2]$
\end{tabular} \\ \hline

$\Spin(12)$ & $\pm \Ga$ & 4 &
\begin{tabular}{c}
$\{(z_1,z_2,-1) \mid z_i^2 = 1\}^{\dag}$\\
$\{(z_1,z_2,\pm 1) \mid z_i^4 = 1\}^{\dag}$
\end{tabular}
\begin{tabular}{c}
$[1]$ \\ $[2]$
\end{tabular} \\ \hline

$\Sp(2)$ & $-\id$ & 2 &
\begin{tabular}{c}
$\{(z_1,z_2,1) \mid z_i^2 = 1\}^{\dag}$\\
$\{(z_1,z_2,z_3) \mid z_i^2 = 1\}^{\dag}$
\end{tabular}
\begin{tabular}{c}
$[1]$ \\ $[1]$
\end{tabular} \\ \hline

$E_6$ & $e^{\pm 2 \pi i / 3} \cdot \id$ & 2,6 &
\begin{tabular}{c}
$\{(z_1,z_2,z_3) \mid z_i^2=1\}^{\dag}$ \\
$\{(z_1,z_2,\pm 1) \mid z_i^6 = 1\}^{\dag}$
\end{tabular}
\begin{tabular}{c}
$[1]$ \\ $[1]$
\end{tabular} \\ \hline

$E_7$ & $-\id$ & 3,6 &
\begin{tabular}{c}
$\{(z_1,z_2,z_3) \mid z_i^3 = 1\}^{\dag}$\\ 
$\{(z_1,z_2,z_3) \mid z_{1,2}^6 = 1\,, z_3^3 = 1\}^{\dag}$
\end{tabular}
\begin{tabular}{c}
$[1]$ \\ $[1]$
\end{tabular} \\ \hline

\end{tabular}
\label{tab:ResultSummaryAlmostCommuting}
\end{center}
\end{table}

\subsection{Lie algebra gradations}
\label{sec:LieAlgebraGradations}
We have previously remarked, and it is also evident from the tables \ref{tab:ResultSummaryCommuting}-\ref{tab:ResultSummaryAlmostCommuting}, that all eigenvalues of the adjoint action of the three holonomies are roots of unity. This implies that all rank zero triples $(U_1,U_2,U_3)$ define $\ZZ_r^3$ gradations of the corresponding Lie algebra for some integer $r$ depending on $G$, $m$ and the order $k$ of the triple. The structure of these gradings are present in the results but require some further explanation.

In general, a gradation of a Lie algebra $\g$ by an abelian group\footnote{In the most general case we can also allow for $\Ga$ to be an abelian semigroup.} $\Ga$ is a decomposition of $\g$ into a direct sum of vector spaces according to
\be
\label{eqn:GradedLieAlgebra}
\g = \underset{p \in \Ga}{\bigoplus} \, \g_p
\ee
such that the Lie bracket satisfies
\be
\label{eqn:BracketCondition}
[\g_p,\g_q] \subseteq \g_{(p+q)}
\ee
where $+$ denotes the group multiplication of $\Ga$. Several subspaces $\g_p$ may be empty and the dimensions of the non-zero subspaces need not be related, provided of course that the total dimension is that of the Lie algebra.

In the case of the eigenvalue vectors $\vec{\la}$, we take the group to be $\Ga = \ZZ_r^3$, $r$ being the highest order of the roots of unity that appear in the spectrum for a particular triple $(U_1,U_2,U_3)$. The group element $p \in \Ga$ is represented by the vector $\vec{\la}$ of eigenvalues and the group multiplication operation $+$ acts independently in each component of the vector by multiplication of the eigenvalues $\la_i$. With this definition of the group multiplication, the Lie bracket satisfies the required property (\ref{eqn:BracketCondition}), since
\be
\ad{U_i}[T_{\la},T_{\la'}] = \la_i \la'_i [T_{\la},T_{\la'}] \,.
\ee 

Since all the triples we are concerned with have zero rank, i.e. the adjoint action of the triples are non-trivial on all generators of the corresponding Lie algebra, the subspace $\g_{(1,1,1)}$ is empty in all gradations. Equivalently, any two generators $T_{\la}$ and $T_{\la'}$ with $\la + \la' = (1,1,1)$ under the adjoint action of a rank zero triple must commute\footnote{Note that generators with $\la + \la' \neq (1,1,1)$ may still commute since all subspaces $\g_p$ contain the origin.}. We can now examine how the structure of Lie algebra gradations are encoded in the sets of eigenvalue vectors.

\subsubsection{Commuting triples}
We first consider the commuting rank zero triples, i.e. the triples with $m = \id$, whose spectra are tabulated in Table \ref{tab:ResultSummaryCommuting}. The order $k$ of a commuting triple is the order of the holonomies $U_i$ in $G$ and consequently these triples define $\ZZ_k^3$ gradations of the corresponding Lie algebras. Furthermore, due to the $\SL(3,\ZZ)$-invariance of the spectra, the only empty subspace is $\g_{(1,1,1)}$. In the case when $k$ is prime all non-zero subspaces $\g_p$ must have the same dimension, as pointed out in \cite{Kac:1999}. 

For general $k$, all subsets $\{(z_1,z_2,z_3) \mid z_i^j = 1\}^{\dag}$ with $j \mid k$ are compatible with the $\SL(3,\ZZ)$-invariance requirement. Subspaces corresponding to subsets where $j$ is prime are of course of equal dimension according to the argument in the previous paragraph. However, the dimension of the Lie algebra uniquely determines the degeneracy of the eigenvalue vectors, or equivalently the dimensions of the subspaces $\g_p$ corresponding to each $p \in \ZZ_k^3$. Thus, we find that, due to $\SL(3,\ZZ)$-invariance, the spectra in Table \ref{tab:ResultSummaryCommuting} are uniquely determined by the order $k$ of the commuting rank zero triple and the dimension of the corresponding Lie algebra.

Consider for example the $k=4$ triple in $G=E_7$; in the corresponding $\ZZ_4^3$ gradation the seven subspaces $\g_p$ with $p \in \{(z_1,z_2,z_3) \mid z_i^2 = 1\}^{\dag}$ have dimension $\mathrm{dim}(\g_p) = 3$, while the remaining subspaces have dimension $\mathrm{dim}(\g_p) = 2$, yielding in total the 133 dimensions of $E_7$. These subspace dimensions are the only ones compatible with the total dimension of the Lie algebra, as remarked in the previous paragraph.

\subsubsection{Almost commuting triples}
In a similar way the almost commuting rank zero triples, corresponding to non-trivial topology $m \neq \id$, define certain gradations. In these cases the order $k$ is not the order of the holonomies as discussed in section \ref{sec:ModuliSpaceConstruction}. However, we note from Table \ref{tab:ResultSummaryAlmostCommuting} that the maximal possible $k$ for each group\footnote{The maximal value of $k$ for the case of $G=\SU(n)$ is $k_{\mathrm{max}}=n$.}, $k_{\mathrm{max}}$, defines the gradations, reflecting the fact that all almost commuting triples in a group have the same spectrum. Thus, they define $\ZZ_{k_{\mathrm{max}}}^3$ gradations of the underlying Lie algebras. The important difference in the case of non-trivial $m$, however, is that this interpretation generically requires additional subspaces, besides $\g_{(1,1,1)}$, to be empty due to the absence of $\SL(3,\ZZ)$-invariance in the spectra. It is therefore not as straightforward to explain, in terms of the $\ZZ_{k_{\mathrm{max}}}^3$ gradations, the result for the $m \neq \id$ case as it was for the case of trivial topology.








%% file: WeakCouplingSpectrum2008.bbl
\begin{thebibliography}{20}
\bibitem{Brink:1977}
L. Brink, J.H. Schwarz, J. Scherk, {\em Supersymmetric Yang-Mills theories}, Nucl.Phys. {\bf B121} (1977) 77

\bibitem{Borel:1953} A. Borel, J.-P. Serre, {\em Sur certain sous-groupes des groupes de Lie compacts}, Comment. Math. Helv {\bf 27} (1953) 128

\bibitem{Borel:1961} A. Borel, {\em Sous-groupes commutatifs et torsion des groupes de Lie compacts connexes}, Tohoku Math. J. {\bf 13} (1961) 216

\bibitem{Witten:1998}
E. Witten, {\em Toroidal compactification without vector structure}, JHEP {\bf 9802} (1998) 006, \href{http://arxiv.org/abs/hep-th/9712028}{\tt arXiv:hep-th/9712028}

\bibitem{Keurentjes:1998} A. Keurentjes, A. Rosly, A. Smilga, {\em Isolated vacua in supersymmetric Yang-Mills theories}, Phys. Rev. {\bf D58} (1998) 081701, \href{http://arXiv.org/abs/hep-th/9805183}{\tt arXiv:hep-th/9805183}

\bibitem{Kac:1999}
V.G. Kac, A.V. Smilga, {\em Vacuum structure in supersymmetric Yang-Mills theories with any gauge group} in Shifman M.A.(ed.): {\em The many faces of the superworld} 185-234, \href{http://arXiv.org/abs/hep-th/9902029}{\tt arXiv:hep-th/9902029}

\bibitem{Keurentjes:1999a} A. Keurentjes, {\em Non-trivial flat connections on the 3-torus I: $G_2$ and the orthogonal groups}, JHEP {\bf 05} (1999) 001, \href{http://arXiv.org/abs/hep-th/9901154}{\tt arXiv:hep-th/9901154}

\bibitem{Keurentjes:1999b} A. Keurentjes, {\em Non-trivial flat connections on the 3-torus II: The exceptional groups $F_4$ and $E_{6,7,8}$}, JHEP {\bf 05} (1999) 014, \href{http://arXiv.org/abs/hep-th/9902186}{\tt arXiv:hep-th/9902186}

\bibitem{Borel:1999}
A. Borel, R. Friedman, J.W. Morgan, {\em Almost commuting elements in compact Lie groups}, \href{http://arxiv.org/abs/math/9907007}{\tt arXiv:math/9907007 [math.GR]}

\bibitem{Henningson:2008}
M. Henningson, N. Wyllard, {\em Zero-energy states of $\cN = 4$ SYM on $T^3$: S-duality and the mapping class group}, JHEP {\bf 04} (2008) 066, \href{http://arXiv.org/abs/0802.0660}{\tt arXiv:0802.0660 [hep-th]}

\bibitem{Witten:2000}
E. Witten, {\em Supersymmetric index in four-dimensional gauge theories}, Adv. Theor. Math. Phys. {\bf 5} (2002) 841, \href{http://arxiv.org/abs/hep-th/0006010}{\tt arXiv:hep-th/0006010}

\bibitem{Henningson:2007a}
M. Henningson, N. Wyllard, {\em Low-energy spectrum of $\cN = 4$ super-Yang-Mills on $T^3$: flat connections, bound states at threshold and S-duality}, JHEP {\bf 06} (2007) 084, \href{http://arXiv.org/abs/hep-th/0703172}{\tt arXiv:hep-th/0703172}

\bibitem{Henningson:2007b}
M. Henningson, N. Wyllard, {\em Bound states in $\cN = 4$ SYM on $T^3$: $\Spin(2n)$ and the exceptional groups}, JHEP {\bf 07} (2007) 001, \href{http://arXiv.org/abs/0706.2803}{\tt arXiv:0706.2803 [hep-th]}

\end{thebibliography}
